\documentclass[prd,twocolumn,showpacs,preprintnumbers,superscriptaddress,floatfix,nofootinbib]{revtex4-1}
\usepackage{graphicx,,booktabs}
\usepackage{amssymb}
\usepackage{amsmath,txfonts}
\usepackage{graphicx}
\usepackage{dcolumn}
\usepackage{color}
\usepackage{bm}
\usepackage[colorlinks,
citecolor=blue,
anchorcolor=green,
menucolor=orange,
linkcolor=red,
filecolor=red,
runcolor=pink,
urlcolor=blue,
frenchlinks=red]{hyperref}
\usepackage{multirow}
\usepackage{slashed}

\begin{document}
	
	\title{\boldmath Prospects for detecting the hidden-strange pentaquarklike state $N^{*}(2080)$ in the $\pi^{-} p\rightarrow\phi n$ reaction}
	\author{Xiao-Yun Wang}
	\email{xywang@lut.edu.cn}
	\affiliation{Department of physics, Lanzhou University of Technology,
		Lanzhou 730050, China}
	\affiliation{Lanzhou Center for Theoretical Physics, Key Laboratory of Theoretical Physics of Gansu Province, Lanzhou University, Lanzhou, Gansu 730000, China}
	
	\author{Hui-Fang Zhou}
	\affiliation{Department of physics, Lanzhou University of Technology,
		Lanzhou 730050, China}
	
	\author{Xiang Liu}
	\email{xiangliu@lzu.edu.cn}
	\affiliation{School of Physical Science and Technology, Lanzhou University, Lanzhou 730000, China}
	\affiliation{Lanzhou Center for Theoretical Physics, Key Laboratory of Theoretical Physics of Gansu Province, Lanzhou University, Lanzhou, Gansu 730000, China}
	\affiliation{Key Laboratory of Quantum Theory and Applications of MoE, Lanzhou University,
		Lanzhou 730000, China}
	\affiliation{MoE Frontiers Science Center for Rare Isotopes, Lanzhou University, Lanzhou 730000, China}
	\affiliation{Research Center for Hadron and CSR Physics, Lanzhou University and Institute of Modern Physics of CAS, Lanzhou 730000, China}

	\begin{abstract}
In this work, the production of the hidden-strange pentaquarklike state $N^{*}(2080)$ via the $\pi^{-} p$ scattering process is studied by the effective Lagrangian approach. Concretely, we consider the $\rho$ meson exchange of $t$-channel and the nucleon exchange of $u$-channel, which are treated as the background terms, and take into account the contribution of the $N^{*}(2080)$ via the $s$-channel as a signal term. By global fitting of the total and differential cross sections of the $\pi^{-} p\rightarrow\phi n$ process, it is shown that the $N^{*}(2080)$ contributes an obvious peak at the threshold energy point of the differential cross section at the forward angle ($\cos\theta = 1$), which provides clues to the detection of the $N^{*}(2080)$ by the $\pi^{-} p$ scattering process. However, due to the limited accuracy of the experimental data, it is an objective limitation for us to determine the properties of the $N^{*}(2080)$ by the $\pi^{-} p$ scattering process. Therefore, more accurate experimental measurements of the $\pi^{-} p\rightarrow\phi n$ reaction are highly desirable, and we have also proposed that correlation measurements can be performed on J-PARC, AMBER, and future HIKE and HIAF meson beam experiments.
	\end{abstract}
	
	\maketitle
	
	\affiliation{Department of physics, Lanzhou University of Technology,
		Lanzhou 730050, China}
	
	\affiliation{Department of  Physics and Institute of Theoretical Physics, Nanjing Normal University,
		Nanjing, Jiangsu 210097, China}
	
	\section{Introduction}
     Since the beginning of the 21st century, with the accumulation of experimental data, more and more new hadronic states have been experimentally observed  \cite{Belle:2003nnu,Klempt:2007cp,BESIII:2013ris,LHCb:2015yax,Chen:2016qju,BESIII:2017bua,LHCb:2019kea,BESIII:2020qkh,LHCb:2020jpq,LHCb:2021vvq,LHCb:2022ogu}. These novel phenomena provide good opportunities to improve our knowledge of hadron spectroscopy and to deepen our understanding of nonperturbative behavior of the strong interaction. 

As an important advance in the study of hadron specstroscopy, the discovery of hidden-charm pentaquarks by the LHCb Collaboration in 2015 \cite{LHCb:2015yax} opens a new window to explore exotic hadronic states. Here, two hidden-charm pentaquarks $P_{c}(4450)$ and $P_{c}(4380)$ were reported in the $\Lambda_{b}^{0}\rightarrow J/\psi K^{-} p$ decay \cite{LHCb:2015yax}. 
With more precise experimental data collected by LHCb, a characteristic mass spectrum was found in the $J/\psi p$ invariant mass spectrum of the $\Lambda_{b}^{0}\rightarrow J/\psi K^{-} p$ decay \cite{LHCb:2019kea}, where the previous $P_{c}(4450)$ splits into two narrow structures, renamed the $P_{c}(4440)$ and $P_{c}(4457)$, and $P_{c}(4312)$ was reported for the first time \cite{LHCb:2019kea}. This observation provides strong evidence for the existence of hidden-charm molecular pentaquarks, as predicted in Refs. \cite{Wu:2010jy,Wang:2011rga,Yang:2011wz,Wu:2012md,Li:2014gra,Chen:2015loa,Karliner:2015ina,Chen:2019asm,Chen:2019bip,He:2019ify,Cheng:2019obk,Meng:2019ilv,Liu:2019tjn,Du:2019pij,Liu:2023wfo}. In the past years, LHCb still has more contributions to explore hidden-charm pentaquarks. As predicted in Ref. \cite{Chen:2016ryt}, there exist possible hidden-charm pentaquarks from $\Sigma_c^{(*)}\bar{D}_s^*$ and $\Xi_c^{(\prime,*)}\bar{D}^*$ interactions. Later, LCHb indeed found the evidence of hidden-charm pentaquark $P_{cs}(4459)$ containing strange quark in the $J/\psi\Lambda$ invariant mass spectrum measured by the $\Xi_{b}^{-}\rightarrow J/\psi K^{-} \Lambda$ process \cite{LHCb:2020jpq}. 
In addition, LCHb discovered the $P_{cs}(4338)$ in the $J/\psi\Lambda$ invariant mass spectrum of the $B^{-}\rightarrow J/\psi \Lambda\bar{p}$ weak decay \cite{LHCb:2022ogu}. The discovery of two hidden-charm pentaquarks with strangeness has induced theorists to conduct extensive discussions on their properties  \cite{Feijoo:2022rxf,Wang:2023ael,Xiao:2019gjd,Wang:2019nvm,Wang:2022nqs,Zhou:2022gra,Wang:2022tib,Wang:2022mxy,Wang:2023iox}. Currently, the heavy flavor pentaquark states are still the focus for hadron spectroscopy studies \cite{Wang:2023aob,Wang:2023mdj,Wang:2022aga,Karliner:2022erb}.
In addition to the experimental search for the hidden-charm pentaquark state via hadronic decay, as done by LHCb, theorists and experimentalists have long been eager to obtain information about the hidden-charm pentaquark state from photoproduction experiments \cite{Duran:2022xag,GlueX:2019mkq,GlueX:2023pev,Gao:2017hya}.
 For example, JLab recently performed a high-precision $J/\psi $ photoproduction experiment near the threshold \cite{Duran:2022xag}, and one of the goals of the recent $J/\psi $ photoproduction measurements performed by the GlueX Collaboration is also to search for hidden-charm pentaquarks \cite{GlueX:2019mkq,GlueX:2023pev}. However, no evidence for the existence of the hidden-charm pentaquark state has yet been obtained in the $J/\psi$ photoproduction experiments, which should be clarified by more precise data in future.

 Due to the existence of hidden-charm pentaquarks as mentioned above, we naturally conjecture the possible hidden-strange pentaquarks, since two systems are similar to each other. There have been some discussions on this subject \cite{He:2003nss,He:2003vi,He:2015yva,Yuan:2010cfa,An:2018vmk,Wu:2011zzm,Zou:2005xy}. For example, the $N^{*}(1535)$ is suggested to have a hidden-strange pentaquark component, so there is a strong coupling with the $\phi N$ channel \cite{An:2009uv,Cao:2009ea}. And then, recently the $N^{*}(2080)$ is considered the hidden-strange
 partner for the hidden-charm molecular pentaquark $P_{c}(4457)$ because its mass is just below the threshold of $K^{*}\Sigma$ \cite{Lin:2018kcc,Xie:2010yk,He:2017aps,Ben:2023uev}\footnote{As noted in the latest version of the Particle Date Group (PDG) review \cite{Workman:2022ynf}, the two-star state $N^{*}(2080)$ is split into a three-star state $N(1875)$ and a two-star state $N(2120)$, and since the corresponding threshold of the hidden-charm hadronic molecular state $K^{*}\Sigma$ is 2086 MeV, we still use the name the $N^{*}(2080)$ in this paper, with a mass of 2080 MeV and a decay width of 99 MeV \cite{Wu:2023ywu,Lin:2018kcc}.}. Considering the $N^{*}(2080)$ to be as a hidden-strange pentaquarklike state, in Ref. \cite{Wu:2023ywu} the authors investigated 
the contribution of the $N^{*}(2080)$ to the $s$-channel of the photoproduction process $\gamma p\rightarrow \phi p $. This motivation is similar to that for the search for hidden-charm pentaquarks via the $J/\psi$ photoproduction experiment.

Focusing on the production of the hidden-strange pentaquarklike state $N^{*}(2080)$, we can borrow some research experiences from the production of light flavor hadrons. By checking the PDG \cite{Workman:2022ynf}, we may find that most of the light flavor hadrons can be produced by reaction processes induced by the $K$ or $\pi$ meson beam. 
In this work, we propose to 
study the production of the hidden-strange pentaquarklike state $N^{*}(2080)$ by the $\pi^{-} p$ scattering. In fact, the $\pi^{-} p$ scattering processes are an ideal platform to search for new hadronic states as indicated in Refs. \cite{Cheng:2016ddp,Wang:2017qcw,Wang:2019zaw,Xie:2013db}. We use the effective Lagrangian approach to calculate and analyze the production rate of the hidden-strange pentaquarklike state $N^{*}(2080)$ by the $\pi^{-} p\rightarrow\phi n$ reaction, where we introduce the $t$-channel $\rho$ exchange and the $u$-channel nucleon exchange as background contributions in the study. Through this analysis, the feasibility of searching for the hidden-strange pentaquarklike state $N^{*}(2080)$ by the $\pi^{-} p$ scattering process is discussed, which is also a possible new task for future meson beam experiments.

     This paper is organized as follows. After the introduction, the Lagrangians and amplitudes used in this work are given in Sec. \ref{sec2}. The numerical results of the total section and differential sections are shown in Sec. \ref{sec3}, and a brief summary of this work is given in Sec. \ref{sec4}.

	\section{The production of the $N^*(2080)$ via the $\pi^{-} p\rightarrow\phi n$ reaction}
    \label{sec2}

\begin{figure}[htbp]
		\centering
		\includegraphics[scale=0.4]{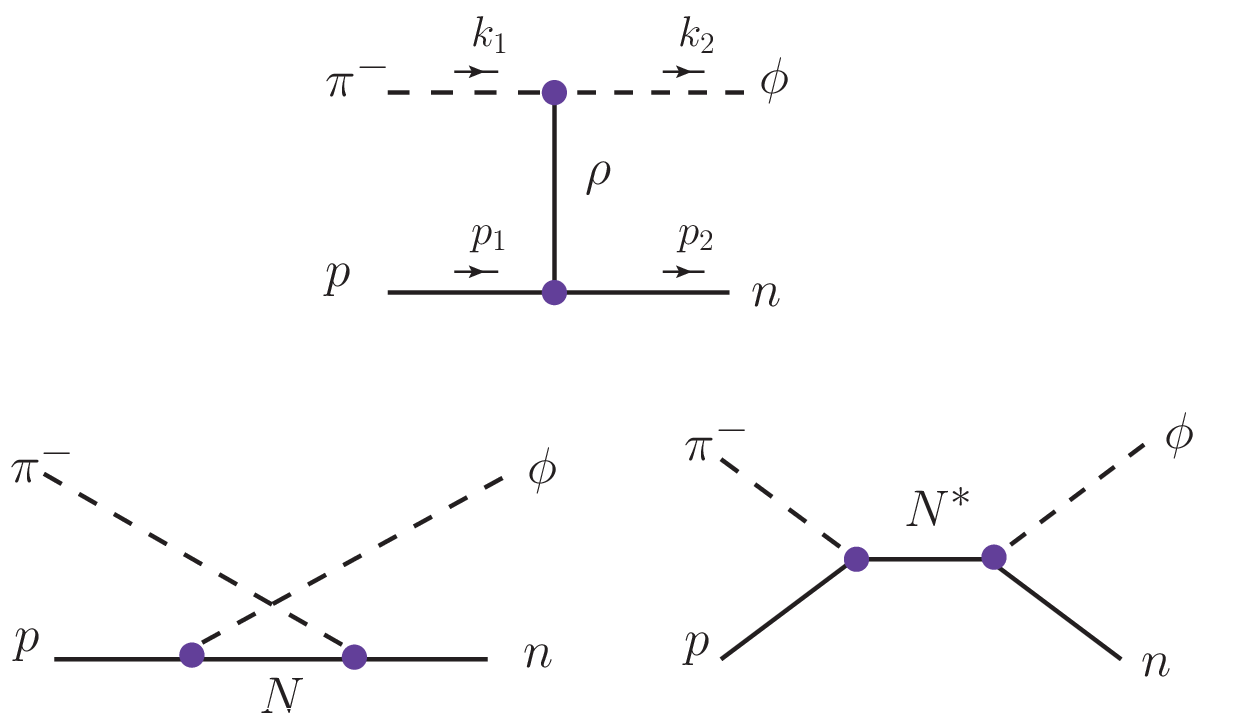}
		\caption{Feynman diagrams for the $\pi^{-} p\rightarrow\phi n$ reaction. The top figure is the $t$-channel tree diagram, while the bottom left and bottom right are the tree diagrams of $u$-channel and $s$-channel, respectively.}
		\label{fig:pip}
	\end{figure}

	The involved tree-level Feynman diagrams of the $\pi^{-} p\rightarrow\phi n$ reaction are shown in Fig. \ref{fig:pip}, which include $t$-channel $\rho(770)(\equiv \rho$) exchange, $u$-channel nucleon exchange and $s$-channel $N^{*}(2080)$$(\equiv N^{*}$) exchange.  
	
	For the $t$-channel $\rho$ exchange  and $u$-channel nucleon exchange, the corresponding Lagrangians are  \cite{Wu:2023ywu,Wang:2019dsi,Ryu:2012tw,Lu:2015pva,Wang:2019krd,Kim:2016cxr,Clymton:2022qlr}
	
	\begin{eqnarray}
		\mathcal{L}_{\pi \rho \phi} &=&\frac{g_{\pi \rho \phi }}{m_{\phi}}\epsilon^{\mu \nu \alpha \beta}\partial_{\mu}\phi_{\nu}\partial_{\alpha}{\bm\rho}_{\beta}\cdot\bm{\pi},\\
		\mathcal{L}_{\rho NN} &=&-g_{\rho NN }\bar{N}\left(\gamma_{\mu}-\frac{\kappa_{\rho NN}}{2m_{N}}\sigma_{\mu \nu}\partial ^{\nu }\right)\bm{{\tau}}\cdot\bm{\rho}^{\mu}{N},
	\end{eqnarray}
	\begin{eqnarray}
\mathcal{L}_{\pi NN} &=&-ig_{\pi NN }\bar{N}\gamma_{5}\bm{\tau}\cdot\bm{\pi}N,\\
    \mathcal{L}_{\phi NN} &=&g_{\phi NN }\bar{N}\left(\gamma_{\mu}-\frac{\kappa_{\phi NN}}{2m_{N}}\sigma_{\mu \nu}\partial ^{\nu }\right)N\phi^{\mu},
	\end{eqnarray}%
 where  $\bm{\pi}$, $\bm\rho$ and $\phi$ are the  $\pi$, $\rho$ and $\phi$ meson fields, respectively. The values of $g_{\pi \rho \phi}=-1.258$, $g_{\rho NN}^{2}/4\pi =0.9$ and $\kappa_{\rho NN}=6.1$ are taken from Ref. \cite{Wu:2023ywu}. $\bm{{\tau}}$ is the Pauli matrix. Additionally, we take $g_{\pi NN}^{2}/4\pi=12.96$  \cite{Wang:2019zaw} and $g_{\phi NN}=-1.47$ \cite{Stoks:1999bz}.
	Since the value of $\kappa_{\phi NN}$ was obtained by fitting the CLAS
	data in our previous work, $\kappa_{\phi NN}=-1.65$ is adopted in this work \cite{Kim:2021adl}.
	
	To gauge the contributions of these diagrams shown in Fig. \ref{fig:pip}, one needs the following Lagrangians for the $s$-channel $N^{*}$ exchange \cite{Wang:2019dsi,Lu:2015fva}. In general, analogous to the $P_{c}$ pentaquark state, the $J^P$ quantum number of the $N^{*}(2080)$ state can be $\frac{3}{2}^{-}$ or $\frac{1}{2}^{-}$. Therefore, we should consider the $N^{*}$ states with two quantum numbers in our calculations.
		
	\begin{eqnarray}
		\mathcal{L}_{\pi N N^{*}}^{3/2^{-}} =\frac{g_{\pi NN^{*}}^{3/2^{-}}}{m_{\pi}}\bar{N}\gamma_{5}\bm{\tau}\cdot\partial_{\mu}\bm{\pi}N^{*\mu}+\mathrm{H.c.},
	\end{eqnarray}
	
	\begin{equation}
		\begin{split}
				\mathcal{L}_{N^{*}\phi N}^{3/2^{-}} =&\frac{-ig_{N^{*}\phi N}^{3/2^{-}}}{2m_{N}}\bar{N}\gamma_{\nu}\phi^{\mu\nu}N^{*}_{\mu}-\frac{g_{2}}{(2m_{N})^{2}}\\
			&\times\partial_{\nu}\bar{N}\phi^{\mu\nu}N^{*}_\mu+\frac{g_{3}}{(2m_{N})^{2}}\bar{N}
			\partial_{\nu}\phi^{\mu\nu}N^{*}_\mu+\mathrm{H.c.},
		\end{split}
	\end{equation}
	
	\begin{eqnarray}
		\mathcal{L}_{\pi N N^{*}}^{1/2^{-}} =g_{\pi NN^{*}}^{1/2^{-}}\bar{N}\bm{\tau}\cdot\bm{\pi}N^{*}+\mathrm{H.c.},
	\end{eqnarray}
	
	\begin{eqnarray}
		\mathcal{L}_{N^{*}\phi N}^{1/2^{-}} =g_{\pi NN^{*}}^{1/2^{-}}\bar{N}\gamma_{5}\gamma_{\mu}N^{*}\phi^{\mu}+\mathrm{H.c.}.
	\end{eqnarray}
	
	 For the $t$-channel $\rho$ meson exchange \cite{Wang:2015xwa}, the general form factor $F_{t}(q_{\rho})$ is given by
 	\begin{eqnarray}
		F_{t}(q_{\rho})&=&\frac{\Lambda _{t}^{2}-m^{2}_{\rho}}{\Lambda _{t}^{2}-q^{2}_{\rho}}.\label{from1}
	\end{eqnarray}
 For the $s$-channel and $u$-channel with intermediate nucleon exchanges,
	we use the general form factor to describe the structure effect of the interaction vertex \cite{Wang:2017qcw,Wang:2023lia,Wang:2015kia}, i.e.,

\begin{eqnarray}
	F_{s/u}(q_N)=\frac{\Lambda_{s/u}^4}{\Lambda_{s/u}^4+(q_N^2-m_N^2)^2}.
\end{eqnarray}
The values of $g_{N^{*}\pi N}$ and $g_{N^{*}\phi N}$ can be determined from the partial decay widths \cite{Liu:2021ojf}, where the involved expressions of these decay widths are
\begin{eqnarray}
	\Gamma_{N^{*}\rightarrow\pi N}^{3/2^{-}}& =\frac{g_{\pi NN^{*}}^{2}(E_{N}-m_{N})}{4\pi m_{\pi}^{2}m_{N^{*}}}|\vec{p}_{N}^{\mathrm{c.m.}}|^{3},
\end{eqnarray}%
and
\begin{eqnarray}
	\begin{split}
		\Gamma_{N^{*}\rightarrow\phi N}^{3/2^-}&=g_{N^{*}\phi N}^2\frac{|\vec{p}_{N}^{\mathrm{c.m.}}|}{192\pi m_{N^{*}}^4m^2_N}[(3m_{N^{*}}^6-m_{N^{*}}^4(m_{\phi}^2+5m_{N}^2)\\
		&\quad+12m_{N^{*}}^3m_{\phi}^2m_{N}+m_{N^{*}}^2(m_{N}^4-m_{\phi}^4)-(m_{\phi}^2-m_{N}^2)^3],
	\end{split}
\end{eqnarray}%

\begin{eqnarray}
	\Gamma_{N^{*}\rightarrow\pi N}^{1/2^{-}}& =\frac{3g_{\pi NN^{*}}^{2}(E_{N}+m_{N})}{4\pi m_{N^{*}}}|\vec{p}_{N}^{\mathrm{c.m.}}|,
\end{eqnarray}%

\begin{eqnarray}
	\begin{split}
		\Gamma_{N^{*}\rightarrow\phi N}^{1/2^-}&=g_{N^{*}\phi N}^2\frac{|\vec{p}_{N}^{\mathrm{c.m.}}|}{8\pi m_{N^{*}}^4m_{\phi}^2}[(m_{N}+m_{N^{*}})^{2}-m_{\phi}^2]
		\\
		&\quad[(m_{N}-m_{N^{*}})^{2}+2m_{\phi}^2],
	\end{split}
\end{eqnarray}%
with
\begin{eqnarray}
	|\vec{p}_{N}^{\mathrm{c.m.}}|& =\frac{\lambda(m_{N^{*}}^2,m_1^2,m_N^2)}{2m_{N^{*}}},
 \label{PN}
\end{eqnarray}%
and
\begin{eqnarray}
	E_{N} =\sqrt{|\vec{p}_{N}^{\mathrm{~c.m.}}|^{2}+m_{N}^{2}}.
\end{eqnarray}
Here, $\lambda $ denotes the K\"{a}llen function with $\lambda (x,y,z)\equiv
\sqrt{(x-y-z)^{2}-4yz}$. Furthermore, for $N^{*}(2080)$ decays to $\pi N$ or $\phi N$, the corresponding $m_1$ in Eq. \ref{PN} takes the mass of $\pi$ or $\phi$ mesons, respectively. At present, there are no experimental measurements of the partial widths of the $N^{*}(2080)$ decays to $\pi N$ and $\phi N$, so we shall determine the coupling constants by fitting them as free parameters in later calculations.

	With the above preparation, the amplitude of the $\pi^{-} p\rightarrow\phi n$ reaction can be
	written as%
	
	\begin{eqnarray}
		-i\mathcal{M} &=& \epsilon^{\mu}(k_{2})\bar{u}(p_{2})\left(\mathcal{A}_{t}^{\rho}+\mathcal{A}_{u}^{N}+\mathcal{A}_{s}^{N^{*}}\right) u_(p_{1}),
	\end{eqnarray}%
	where $\epsilon_{\mu}$ is the polarization vector of the $\phi$ meson. $\bar{u}$ or $u$ denotes the Dirac spinor of the nucleon.
	The reduced amplitudes $\mathcal{A}_{t}^{\rho}$, $\mathcal{A}_{s}^{N^{*}}$ and $\mathcal{A}_{u}^{N}$ for $t$-channel, $s$-channel and $u$-channel contributions read as
	\begin{eqnarray}
	\mathcal{A}_{t}^{\rho} &=&-\sqrt{2}\frac{g_{\pi \rho \phi }}{m_{\phi}}g_{\rho NN }F_{t}(q^{2}_{t})\frac{\mathcal{P}^{\nu \xi}}{t-m_{\rho}^{2}}\epsilon_{\mu \nu \alpha \beta}k_{2}^\alpha \notag \\
&&\times(k_{2}-k_{1})^{\beta}\left[\gamma_{\xi}+\frac{\kappa_{\rho NN}}{4m_{N}}(\gamma_{\xi}q\mkern-8mu/_{t}-q\mkern-8mu/_{t}\gamma_{\xi})\right],
	\end{eqnarray}%
	
	\begin{eqnarray}
		\mathcal{A}_{u}^{N}&=&-i\sqrt{2}g_{\pi NN }g_{\phi NN }F_{u}(q_{N})\left[\gamma_{\mu}+\frac{\kappa_{\phi NN}}{4m_{N}}(\gamma_{\mu}k\mkern-8mu/_{2}-k\mkern-8mu/_{2}\gamma_{\mu})\right] \notag \\
		&&\times\frac{({ q\mkern-8mu/_{N}}+m_{N})}{u-m_{N}^{2}}\gamma_{5},
	\end{eqnarray}

\begin{eqnarray}
	\mathcal{A}_{s}^{N^{*}(3/2^{-})} &=&\sqrt{2}\frac{g_{\pi NN^{*}}^{3/2^{-}}}{m_{\pi}}\frac{-ig_{N^{*}\phi N}^{3/2^{-}}}{2m_{N}}F_{s}(q_{s}^{2})\gamma^{\sigma}(k_{2\beta}g_{\mu\sigma}\notag \\
	&&-k_{2\sigma}g_{\mu\beta}) \frac{(q\mkern-8mu/_{s}+m_{N^{*}})}{s-m_{N^{*}}^{2}+im_{N^{*}}\Gamma_{N^{*}}}\Delta^{\beta\alpha}k_{1\alpha}\gamma_{5},
\end{eqnarray}

\begin{eqnarray}
	\mathcal{A}_{s}^{N^{*}(1/2^{-})} &=\sqrt{2}g_{\pi NN^{*}}^{1/2^{-}}g_{N^{*}\phi N}^{1/2^{-}}F_{s}(q_{s}^{2})\gamma_{5}\gamma_{\mu} \frac{(q\mkern-8mu/_{s}+m_{N^{*}})}{s-m_{N^{*}}^{2}+im_{N^{*}}\Gamma_{N^{*}}},
\end{eqnarray}

	with
  \begin{eqnarray}
     \mathcal{P}^{\nu \xi}=i\left(g^{\nu \xi}+q^{\nu}_{\rho}q^{\xi}_{\rho}/m^{2}_{\rho}\right),
  \end{eqnarray}

	\begin{eqnarray}
	\Delta^{\beta\alpha}&=&-g^{\beta\alpha}+\frac{1}{3}\gamma^{\beta}\gamma^{\alpha} \nonumber\\
	&&+\frac{1}{3m_{N^{*}}}(\gamma^{\beta}q^{\alpha}-\gamma^{\alpha}q^{\beta})+\frac{2}{3m_{N^{*}}^{2}}q^{\beta}q^{\alpha}.
	\end{eqnarray}
     where $s=(k_{1}+p_{1})^{2}$, $t=(k_{1}-k_{2})^{2}$ and $u=(p_{2}+k_{1})^{2}$ are the Mandelstam variables.

	\section{Numerical results}
	\label{sec3}
	
	For the $\pi^{-} p\rightarrow\phi n$ reaction, the differential cross section in the center-of-mass (c.m.) frame is written as
	\begin{equation}
		\frac{d\sigma }{d\cos \theta }=\frac{1}{32\pi s}\frac{\left\vert \vec{k}%
			_{2}^{{~\mathrm{c.m.}}}\right\vert }{\left\vert \vec{k}_{1}^{{~\mathrm{c.m.}}%
			}\right\vert }\left( \frac{1}{2}\sum\limits_{\lambda }\left\vert \mathcal{M}%
		\right\vert ^{2}\right),
	\end{equation}%
	where $\theta $ denotes the angle of the outgoing
	$\phi$ meson relative to $\pi$ beam direction in the c.m. frame. $\vec{k%
	}_{1}^{{~\mathrm{c.m.}}}$ and $\vec{k}_{2}^{{~\mathrm{c.m.}}}$ are the
	three-momenta of the initial $\pi$ beam and the final $\phi$, respectively.
 
\subsection{Production of the $N^{*}(2080)$ with $J^{P}=\frac{3}{2}^{-}$}

 We can analyze the $\pi^{-} p\rightarrow\phi n$ experimental datas based on the theoretical model constructed above. To reduce the number of free parameters, we set $\Lambda_{s}$=$\Lambda_{u}$ and define $g_{N^{*}}^{3/2^{-}}\equiv g_{\pi NN^{*}}^{3/2^{-}}g_{N^{*}\phi N}^{3/2^{-}}$, which is the product of two coupling constants involved in the amplitude for the $N^{*}$ with $J^{P}=\frac{3}{2}^{-}$. The experimental data for the total cross section and the differential cross section when $\cos \theta$ = 1 lead us to use the $\chi^{2}$ fitting algorithm to determine the values of the free parameters. The fitting parameters involved are shown in Table \ref{tab:table1}, with a reduced value of  $\chi^{2}/ d.o.f.=1.05$, indicating that the experimental data of $\pi^{-} p\rightarrow\phi n$ can be well explained by our model. 

\renewcommand\arraystretch{1.5} 
 \begin{table}[htp] \small
	 	\caption{\label{tab:table1} 
	 		{Fitted values of the free parameters by fitting the experimental data in Refs. \cite{Courant:1977rk,Dahl:1967pg} for the case of $N^{*}$ with $J^{P}=\frac{3}{2}^{-}$.}}
	 	\begin{ruledtabular}
	 		\begin{tabular}{ccccc}
	 			&
	 			$\Lambda_{t}$ (GeV)  &$\Lambda_{s/u} $ (GeV)   & $g_{N^{*}}^{3/2^{-}} $ & $\chi^{2}/ d.o.f.$ \\ \hline
	 			Values &1.27$\pm$0.03&0.49$\pm$0.02&0.02$\pm$0.01& 1.05\\ 
	 		\end{tabular}
	 	\end{ruledtabular}
	 \end{table}
 
	\begin{figure}
		\centering
		\includegraphics[scale=0.43]{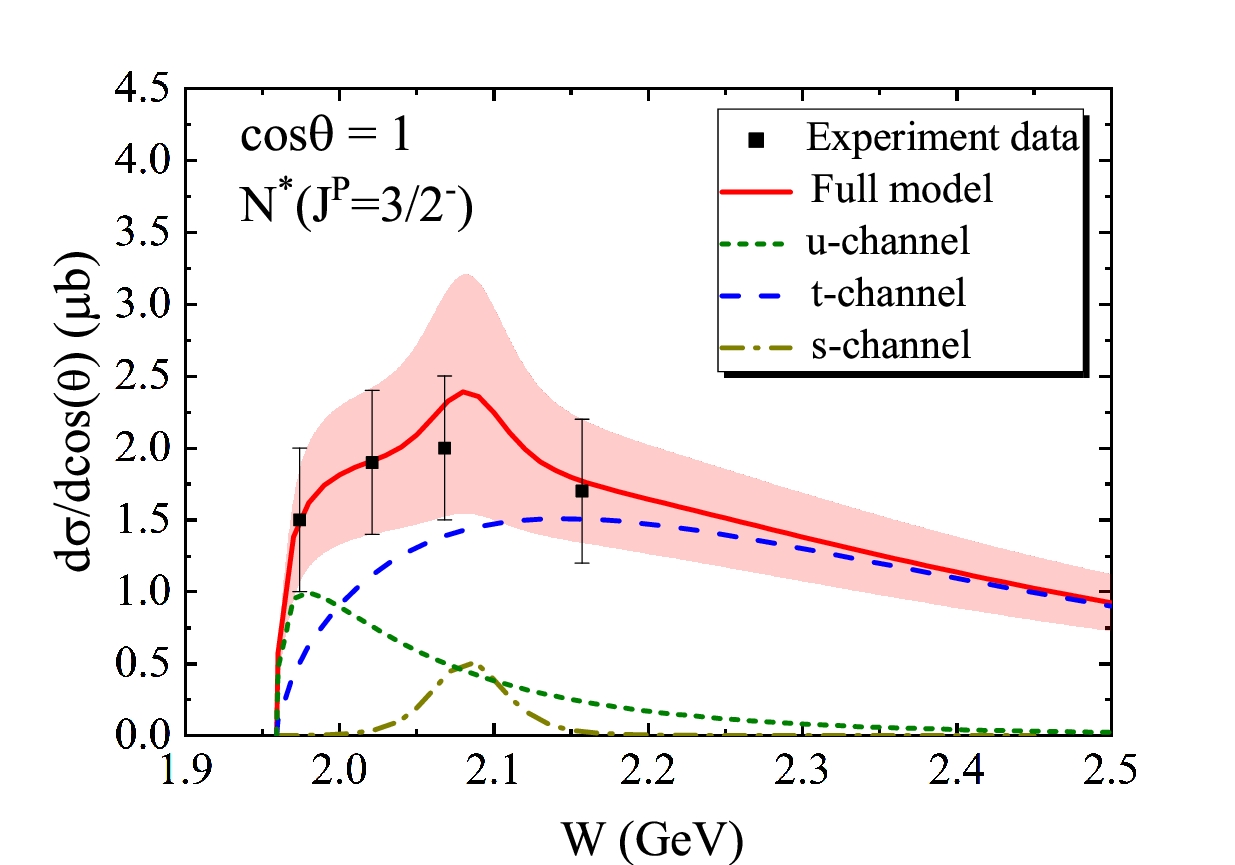}
		\caption{(Color online.) The differential cross section of the $\pi^{-} p\rightarrow\phi n$ reaction varies with different c.m. energies when $\cos\theta$ = 1. The experimental data are taken from Ref. \cite{Courant:1977rk}. The band stands for the error bar of the three fitting parameters in Table \ref{tab:table1}.  The solid (red), short-dashed (green), dashed (blue), dashed-dotted (dark yellow) lines are for the full model, the $u$-channel, the $t$-channel and the $s$-channel, respectively. Here, the spin-parity quantum number of the $N^{*}(2080)$ is $\frac{3}{2}^{-}$.}
		\label{fig:dcsFA}
	\end{figure}

	 \begin{figure}
		\centering
		\includegraphics[scale=0.43]{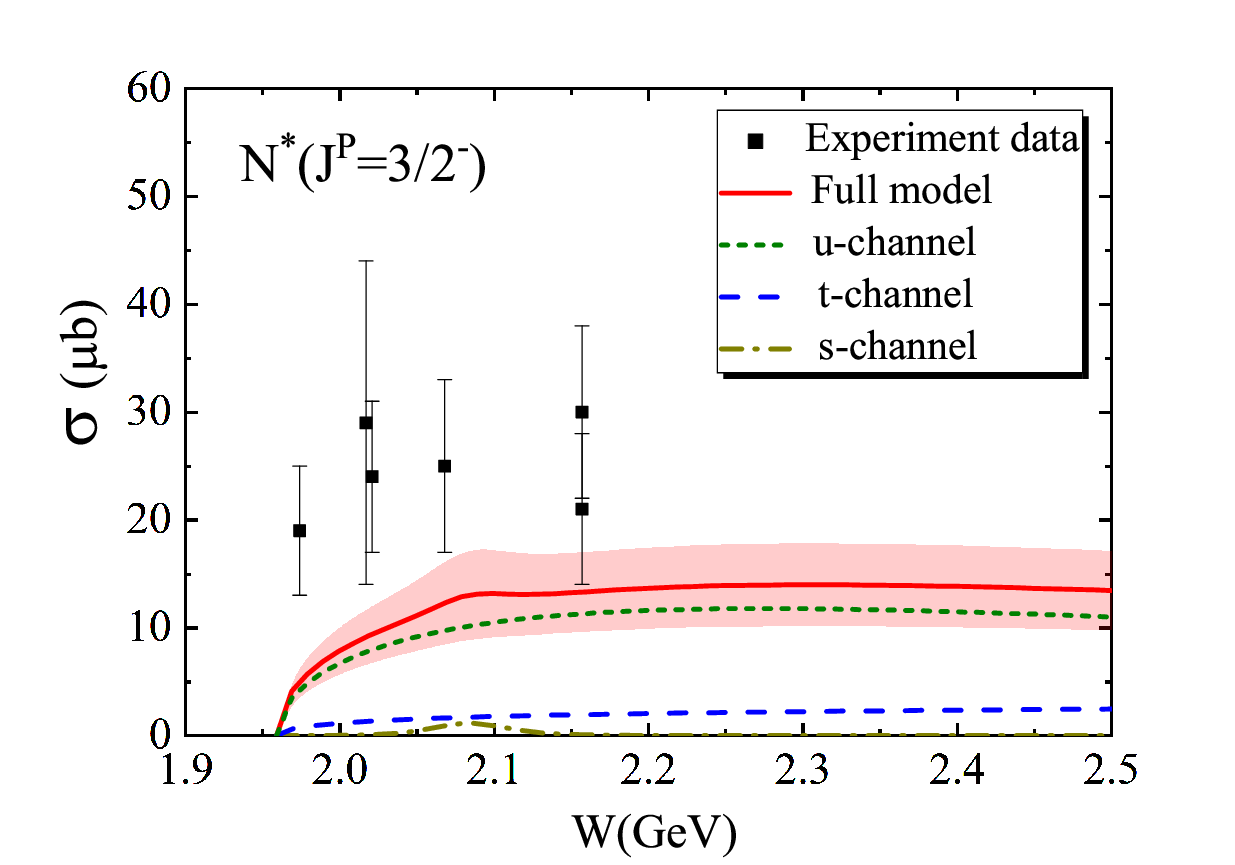}
		\caption{(Color online.) The total cross section for the reaction of $\pi^{-} p\rightarrow\phi n$. The experimental data are taken from Refs. \cite{Courant:1977rk,Dahl:1967pg}. The band stands for the error bar of the three fitting parameters in Table \ref{tab:table1}.   Here, the notation is the same as that in Fig. \ref{fig:dcsFA}.}
		\label{fig:tcsA}
	\end{figure}
	
	 In Fig. \ref{fig:dcsFA}, we present the differential cross section of the $\pi^{-} p\rightarrow\phi n$ reaction as a function of c.m. energies when $\cos\theta$ = 1 for the case of $N^{*}$ with $J^{P}=\frac{3}{2}^{-}$. As shown in Fig. \ref{fig:dcsFA}, the results from the full contribution of the $t$-channel, $u$-channel, and $s$-channel agree well with the experimental data as a whole. Moreover, since these are differential cross sections measured at the forward angle, the curve shapes from the contributions of the three channels are different, which on the one hand can effectively limit the contribution of the $t$-channel and $u$-channel, and on the other hand provides a good measurement object for determining the contribution of the $s$-channel $N^{*}(2080)$. We have also studied the case of the backward differential cross section ($\cos\theta = -1$). The results show that it is difficult to distinguish the contribution of $N^{*}$ from the backward differential cross section because the contribution of the $u$-channel is at least an order of magnitude larger than that of the $t$-channel and $s$-channel. According to the present results, it seems an ideal way to examine the contribution of $N^{*}$ through the differential cross section at the forward angle.

  Fig. \ref{fig:tcsA} shows the total cross section of the $\pi^{-} p\rightarrow\phi n$  reaction. It suggests that the $u$-channel with nucleon exchange plays a dominant role in this process and the contribution of the $t$-channel is relatively small in the total cross section. This situation reminds us that the contribution of  $u$-channel is not negligible, although its contribution at the forward angle is relatively small. Obviously, the total and differential cross section experimental data provide a good constraint and clarification on the $t$-channel and $u$-channel contribution. However, it should be noted that due to the large error of the experimental data and the fact that our theoretical results are generally smaller than the present experimental data, it is not easy to identify the contribution of the $s$-channel $N^{*}$ exchange from the total cross section data.
	  
    \begin{figure}
 	\centering
 	\includegraphics[scale=0.43]{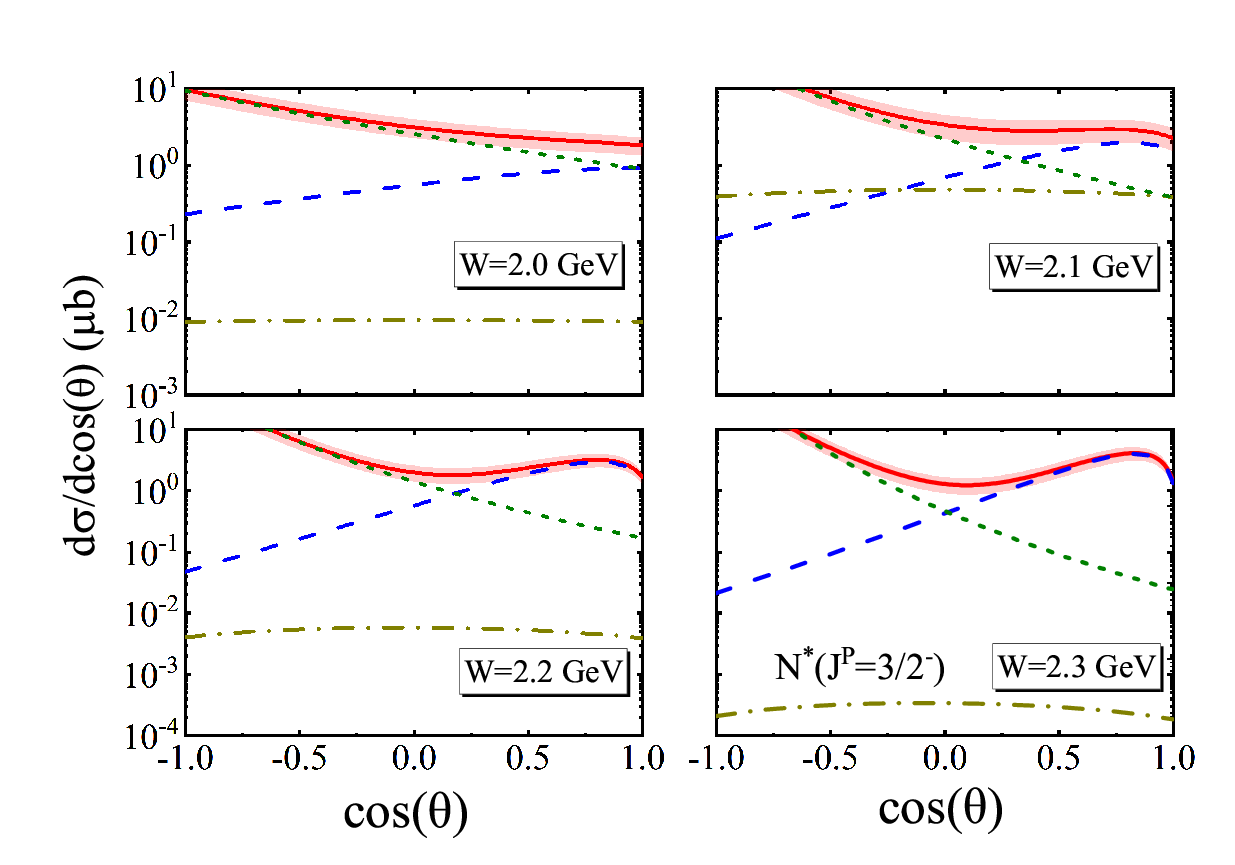}
 	\caption{(Color online.) The differential cross section $d\sigma/d\cos\theta$ of the $\pi^{-} p\rightarrow\phi n$ process as a function of  $\cos\theta$ at different c.m. energies. Here, the notation is the same as that in Fig. \ref{fig:dcsFA}.}
 	\label{fig:dcsA}
    \end{figure}
In Fig. \ref{fig:dcsA}, the differential cross section of the full angle of the $\pi^{-} p\rightarrow\phi n$ reaction at different centre of mass energies is calculated. It can be seen from Fig. \ref{fig:dcsA} that with the increase of energy, the contributions of $t$-channel and $u$-channel become more significant at the forward angle and backward angle respectively, while the overall distribution of $s$-channel contribution is relatively flat. 
	\begin{figure}
 	\centering
 	\includegraphics[scale=0.43]{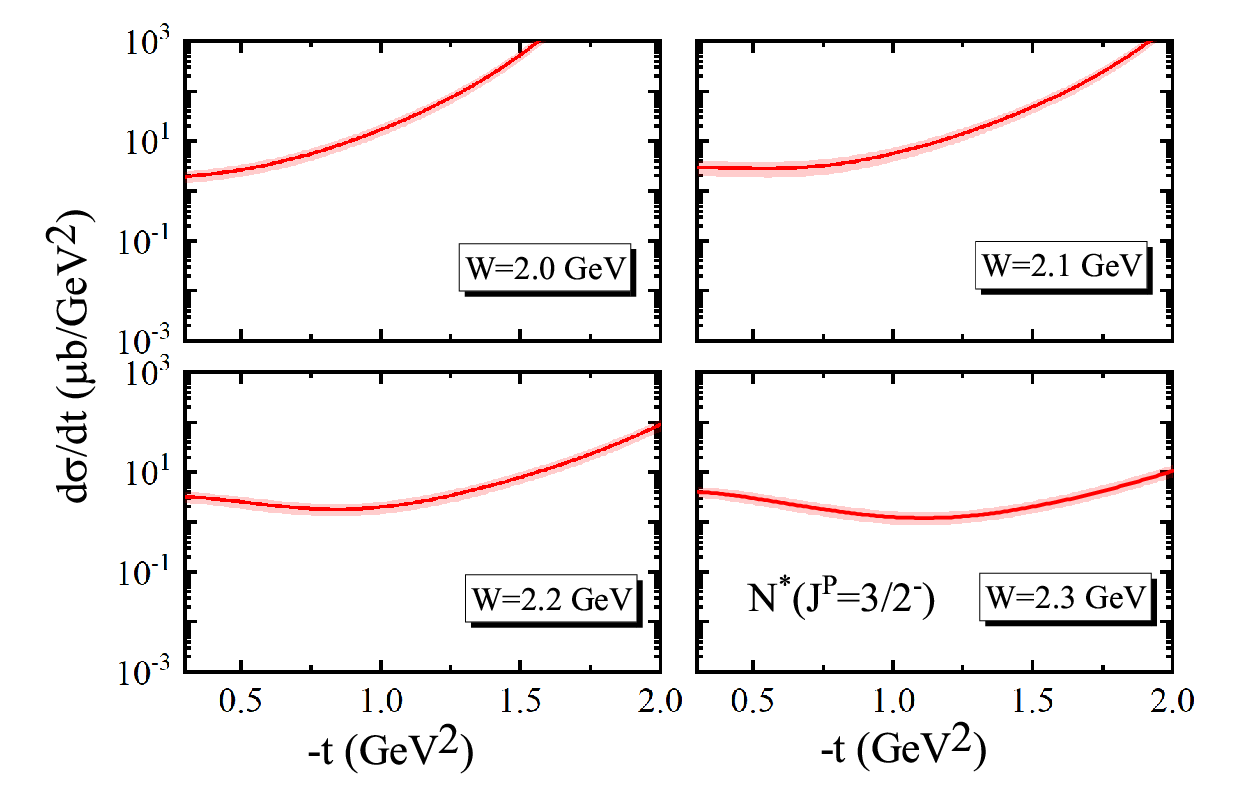}
 	\caption{The $t$-distribution for the $\pi^{-} p\rightarrow\phi n$
reaction at different c.m. energies W = 2.0 GeV, 2.1 GeV, 2.2 GeV and 2.3 GeV.
Here, the notation is the same as in Fig. \ref {fig:dcsFA}.}
 	\label{fig:dcstA}
    \end{figure}
    Fig. \ref{fig:dcstA} shows that the $t$-distributions for the $\pi^{-} p\rightarrow\phi n$ reaction at different energies. It can be seen from the calculation results that the shape of the curve distribution tends to flatten out from the slope as the energy increases. These results may help us distinguish the production mechanism of the $\pi^{-} p\rightarrow\phi n$ reaction.
  
  \subsection{{red}Production of the $N^{*}(2080)$ with $J^{P}=\frac{1}{2}^{-}$}
    
\begin{figure}
    	\centering
    	\includegraphics[scale=0.43]{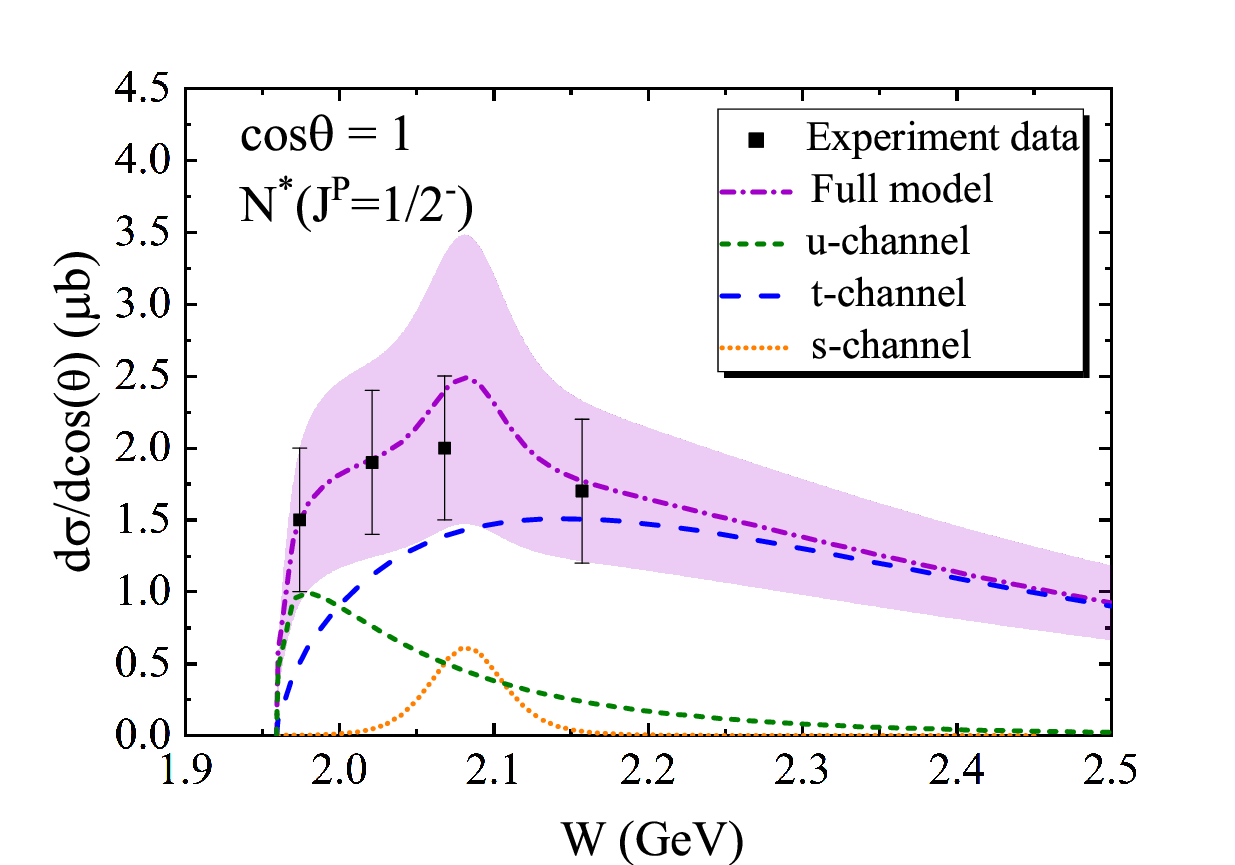}
    	\caption{(Color online.) The differential cross section of the $\pi^{-} p\rightarrow\phi n$ reaction varies with different c.m. energies when $\cos\theta$ = 1. The experimental data are taken from Ref. \cite{Courant:1977rk}. The band stands for the error bar of the three fitting parameters in Table \ref{tab:table2}. The short-dashed dotted (purple), short-dashed (green), dashed (blue) and short-dotted (orange) lines are for the full model, the $u$-channel, the $t$-channel and the $s$-channel, respectively. Here, the spin-parity quantum number of the $N^{*}(2080)$ is $\frac{1}{2}^{-}$.}
    	\label{fig:dcsFB}
    \end{figure}

\begin{figure}
    	\centering
    	\includegraphics[scale=0.43]{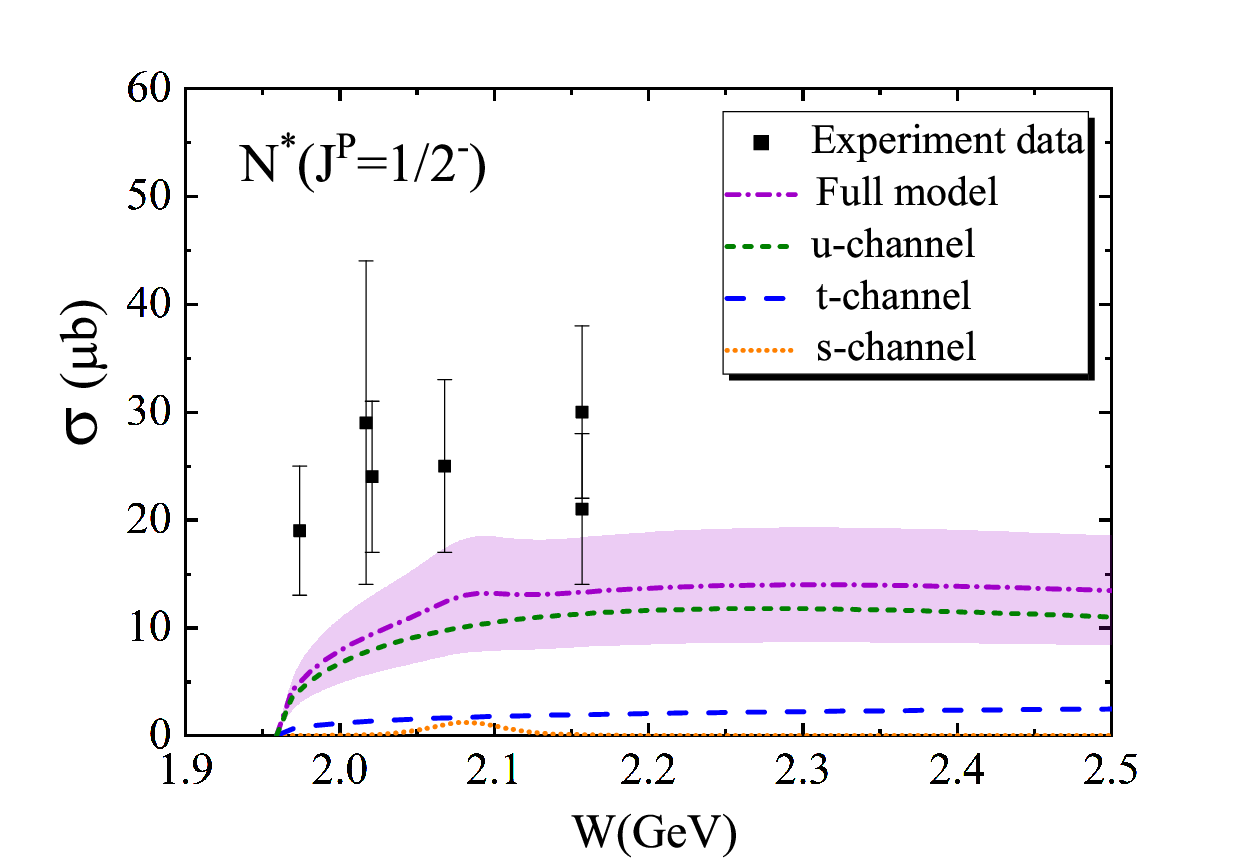}
    	\caption{(Color online.) The total cross section for the reaction of $\pi^{-} p\rightarrow\phi n$. The experimental data are taken from Ref. \cite{Courant:1977rk,Dahl:1967pg}. The band stands for the error bar of the three fitting parameters in Table \ref{tab:table2}. The notation is the same as in Fig. \ref {fig:dcsFB}.}
    	\label{fig:tcsB}
    \end{figure}

\begin{figure}
    	\centering
    	\includegraphics[scale=0.43]{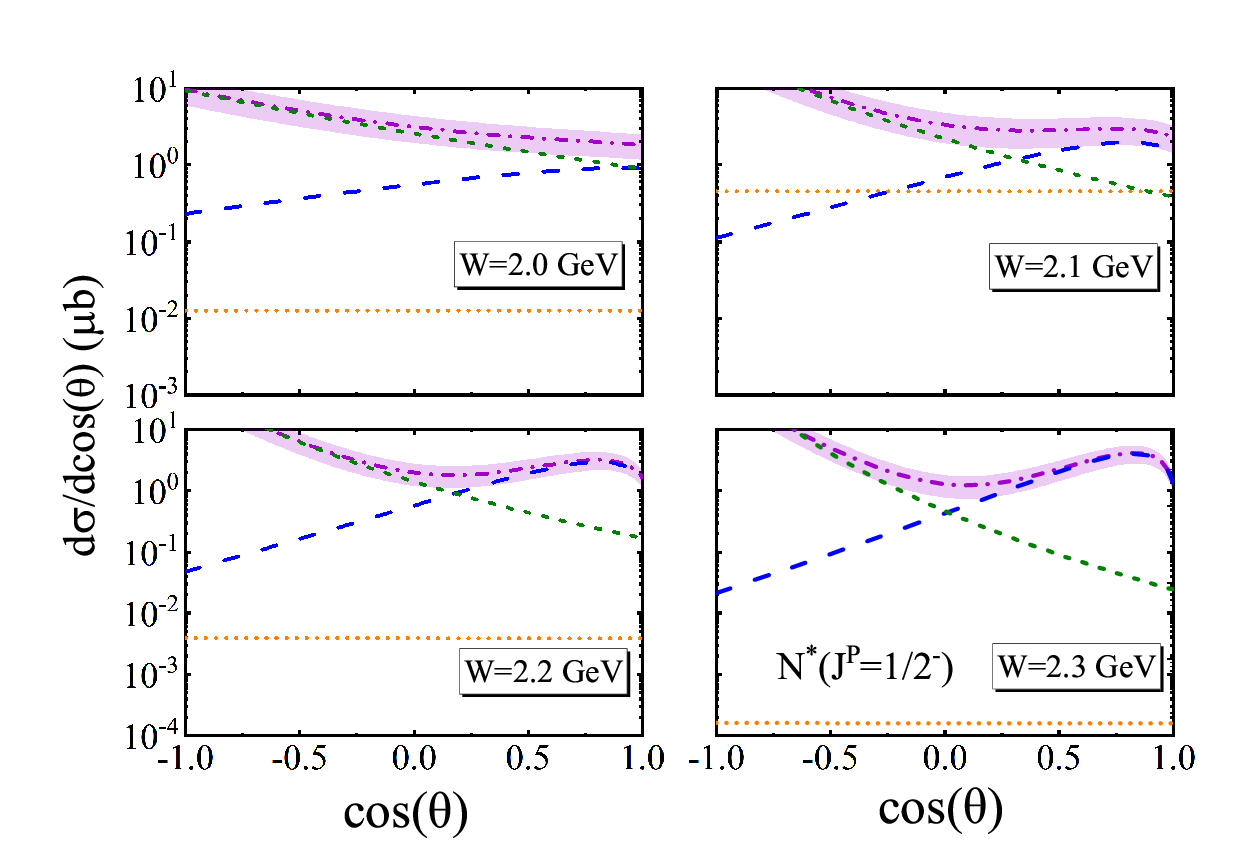}
    	\caption{(Color online.) The differential cross section $d\sigma/d\cos\theta$ of the $\pi^{-} p\rightarrow\phi n$ process as a function of  $\cos\theta$ at different c.m. energies. Here, the notation is the same as that in Fig. \ref{fig:dcsFB}.}
    	\label{fig:dcsB}
    \end{figure}

\begin{figure}
    	\centering
    	\includegraphics[scale=0.43]{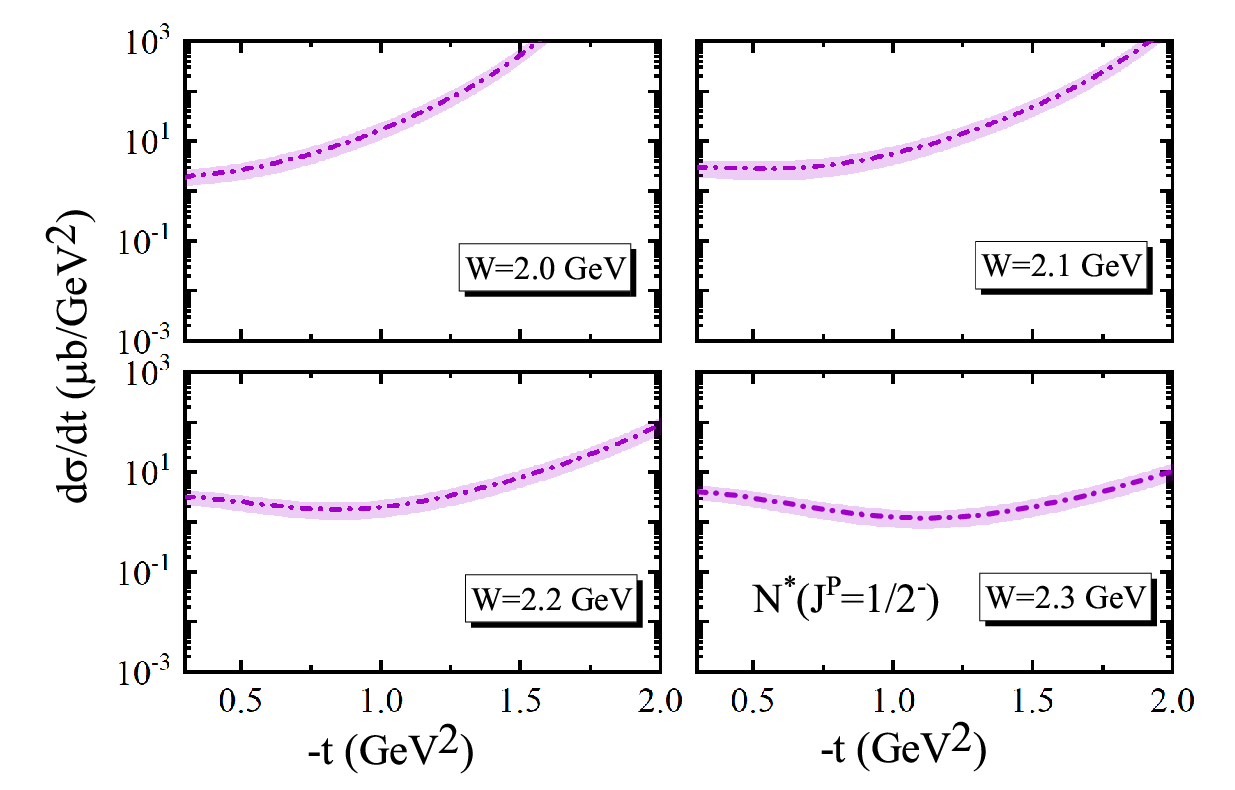}
    	\caption{The $t$-distribution for the $\pi^{-} p\rightarrow\phi n$
    		reaction at different c.m. energies W = 2.0 GeV, 2.1 GeV, 2.2 GeV and 2.3 GeV.
    		Here, the notation is the same as in Fig. \ref {fig:dcsFB}.}
    	\label{fig:dcstB}
    \end{figure}

    By analogy with the calculation of $N^{*}$ with $J^{P}=\frac{3}{2}^{-}$, we also calculate the cross section of $\pi^{-} p\rightarrow\phi n$ for the case of $N^{*}$ with $J^{P}=\frac{1}{2}^{-}$ as shown in Figs. \ref{fig:dcsFB}-\ref{fig:dcstB}. Since the absence of experimental measurements of the partial widths of $N^{*}(J^p=\frac{1}{2}^{-})$ decays to $\pi N$ and $\phi N$, we set the coupling constant $g_{N^{*}}^{1/2^{-}}$($g_{N^{*}}^{1/2^{-}}\equiv g_{\pi NN^{*}}^{1/2^{-}}g_{N^{*}\phi N}^{1/2^{-}}$) as a free parameter determined by fitting. Moreover, we use the same fitting scheme as in Table \ref{tab:table1}, and set the value of cutoff as free parameters for fitting. The fitted parameter values are shown in Table \ref{tab:table2} with a reduced value of $\chi^{2}/ d.o.f.$, one find that the fitting values of the cut off parameters are basically consistent with those in Table \ref{tab:table1}. Fig. \ref{fig:dcsFB} and Fig. \ref{fig:tcsB} show the differential cross section at forward angle and total cross section of the $\pi^{-} p\rightarrow\phi n$ reaction, respectively. One find that these results are very close to those in Fig. \ref{fig:dcsFA} and \ref{fig:tcsA}, and the reason for this result is that the coupling constant g is set to a free parameter, which will play a role in adjusting the size of cross section.
    
    Fig. \ref{fig:dcsB} and Fig. \ref{fig:dcstB} show the full angular differential cross sections and the $t$-distribution of the $\pi^{-} p\rightarrow\phi n$ reaction with different centre of mass energies. For the cross section of the $t$ distribution, there is a slight difference in the shape of the cross section curves corresponding to $N^{*}$ with different quantum numbers. However, since the contributions of the $t$ and $u$ channels are generally larger than the contribution of the $s$ channel with $N^{*}$ exchange, this difference will be covered and it is difficult to distinguish by experiment. The same is true for the distribution of differential cross sections at full angles.

    In short, due to the lack of experimental data and research on the nature of $N^{*}(2080)$, it is difficult for us to accurately give the difference in the influence of $N^{*}$ with different quantum numbers on cross sections. Therefore, our future work will also plan to analyze and calculate the internal properties and decay width of $N^{*}$, which will help to better determine the internal structure and production mechanism of $N^{*}$.
      
    \renewcommand\arraystretch{1.5} 
    \begin{table}[htp] \small
    	\caption{\label{tab:table2} 
    		Fitted values of the free parameters by fitting the experimental data in Refs. \cite{Courant:1977rk,Dahl:1967pg} for the case of $N^{*}$ with $J^{P}=\frac{1}{2}^{-}$.  }
    	\begin{ruledtabular}
    		\begin{tabular}{ccccc}
    			&
    			$\Lambda_{t}$ (GeV)  &$\Lambda_{s/u} $ (GeV)   & $g_{N^{*}}^{1/2^{-}} $ & $\chi^{2}/ d.o.f.$ \\ \hline
    			Values &1.27$\pm$0.04&0.49$\pm$0.03&0.012$\pm$0.005& 0.88\\ 
    		\end{tabular}
    	\end{ruledtabular}
    \end{table}
    
    \subsection{Dalitz process}
\begin{figure}
	\centering
	\includegraphics[scale=0.42]{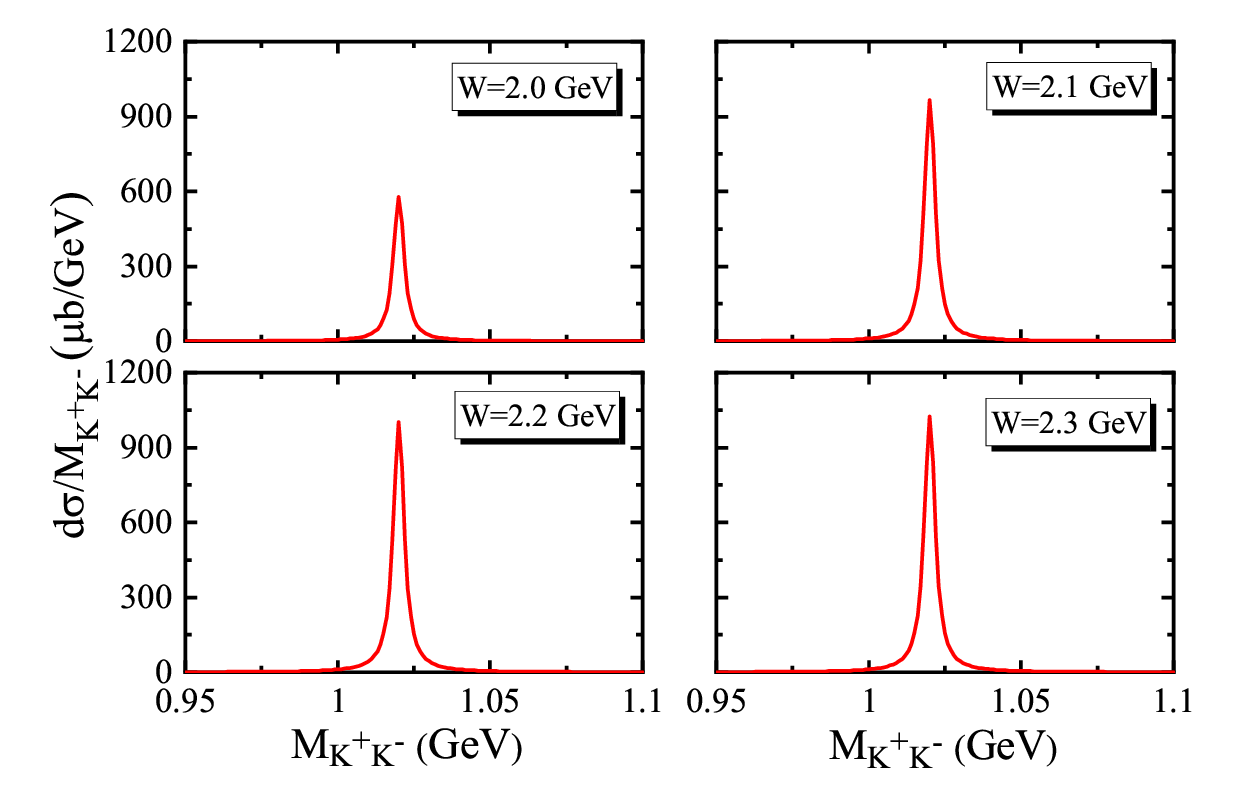}
	\caption{The invariant mass distribution $d\protect\sigma %
		_{\pi^{-}p\rightarrow \protect \phi n \rightarrow \protect K^{+}K^{-}n }/ dM_{\protect K^{+}K^{-} }$ at different c.m. energies W = 2.0 GeV, 2.1 GeV, 2.2 GeV and 2.3 GeV.}
	\label{fig:dalitz}
\end{figure}

    For $\pi^{-} p\rightarrow\phi n$ processes, the $\phi$ meson produced cannot be measured directly experimentally, but needs to be determined by reconstructing the final particles produced by its decay. Considering that the branching ratio of $\phi$ decay to $K^{+}K^{-}$ can reach $49.1\%$ \cite{Workman:2022ynf}, it is necessary to analyze the Dalitz process for $\pi^{-}p\rightarrow \phi n \rightarrow K^{+}K^{-}n$ , which will provide useful information for experimental measurements. According to the above calculation, the total cross section of $\pi^{-} p\rightarrow\phi n$ is very similar in size for the two cases where the quantum number of the $N^{*}(2080)$ is $\frac{3}{2}^{-}$ or $\frac{1}{2}^{-}$. Therefore, we calculate the dalitz process only for the case of $N^{*}$ with $J^{P}=\frac{3}{2}^{-}$. In general, the invariant mass spectrum of the Dalitz process is defined based on the two-body process \cite{Kim:2017nxg}
\begin{equation}
	\frac{d\sigma _{\pi^{-}p\rightarrow \phi n \rightarrow K^{+}K^{-}n }}{dM_{K^{+}K^{-} }}\approx \frac{2M_{\phi}M_{K^{+}K^{-}}}{\pi }\frac{\sigma _{\pi^{-}p\rightarrow \phi n
		}\Gamma _{\phi\rightarrow K^{+}K^{-} }}{(M_{K^{+}K^{-}
		}^{2}-M_{\phi}^{2})^{2}+M_{\phi}^{2}\Gamma
		_{\phi}^{2}},
\end{equation}%
    where $\Gamma _{\phi}=4.249$ MeV and $\Gamma _{\phi\rightarrow K^{+}K^{-} }=2.086$ MeV represent the total width and decay component width of $\phi$, respectively. The results of the calculation are exhibited in Fig. \ref{fig:dalitz}. It is found that the peak values appear at $M_{K^{+}K^{-}} = 1.02$ GeV and the peaks are not less than 577 $\mu b$/GeV when the center of mass energy is 2.0 $\sim$ 2.3 GeV, which means that it is feasible to reconstruct the $\pi^{-} p\rightarrow\phi n$ process by the $\pi^{-}p\rightarrow  K^{+}K^{-}n$ process in experiments. 
    
    The calculated total cross sections for $\pi^{-}p\rightarrow \phi n \rightarrow K^{+}K^{-}n$ and experimental data for the $\pi^{-}p\rightarrow  K^{+}K^{-}n$ process are given in Table \ref{tab:table3}. Furthermore, the percentage of $\pi^{-}p\rightarrow \phi n \rightarrow K^{+}K^{-}n$ reaction cross sections in the $\pi^{-}p\rightarrow  K^{+}K^{-}n$ cross sections is calculated. It is found that the smallest percentage reaches $2.6\%$ in the c.m. energies from 2.15 to 2.46 GeV, which further indicates that the total cross section and the number of events of the $\pi^{-} p\rightarrow\phi n$ process reconstructed and studied by the $\pi^{-}p\rightarrow  K^{+}K^{-}n$ process meets the requirements of the experimental measurement. It is also noted that the ratio of $\sigma(\pi^{-}p\rightarrow \phi n \rightarrow K^{+}K^{-}n)$ to $\sigma(\pi^{-}p\rightarrow K^{+}K^{-}n)$ can reach up to $16.7\%$ at a c.m. energy of 2.15 GeV, of which the contribution of the $s$-channel $N^{*}(2080)$ exchange accounts for $0.13\%$, close to 50 $n b$. Moreover, the cross section of $\sigma(\pi^{-}p\rightarrow \phi n \rightarrow K^{+}K^{-}n)$ corresponding to the $s$-channel $N^{*}(2080)$ exchange can reach 500 $n b$ at a c.m. energy of 2.08 GeV. These results suggest that the number of the $N^{*}(2080)$ produced by the $\pi^{-}p\rightarrow \phi n \rightarrow K^{+}K^{-}n$ process will be considerable, which will be a very important support for future experiments.

\renewcommand\arraystretch{1.5} 

     \begin{table}[htbp]
    \centering
     \caption{\label{tab:table3} 
	 		The total cross section of $\pi^{-}p\rightarrow K^{+}K^{-}n$ and $\pi^{-}p\rightarrow \phi n \rightarrow K^{+}K^{-}n$. Here, the $\sigma_{1}$ represents experimental cross section data of  $\pi^{-}p\rightarrow K^{+}K^{-}n$ \cite{Dahl:1967pg,Boyd:1968pg,Miller:1965pg,Goussu:1966pg}, while $\sigma_{2}$ is the calculated cross section of $\pi^{-}p\rightarrow \phi n \rightarrow K^{+}K^{-}n$.  } 
    \setlength{\tabcolsep}{5mm}{
    \begin{tabular}{c|c|c|c}
    \hline\hline
        $W$ (GeV) & $\sigma_{1}$ $ (\mu b)$ & $\sigma_{2} $ $(\mu b)$ & $(\sigma_{2}/\sigma_{1})$  \\ \hline
        2.15 & 39 & 13.23 & $16.7\%$  \\ \hline
        2.2  & 139 & 13.65 & $4.8\%$  \\ \hline
        2.44 & 84 & 13.70 & $8\%$\\ \hline
        2.46 & 250 & 13.62 & $2.6\%$  \\ \hline\hline
    \end{tabular}}
\end{table}

     \section{SUMMARY}
     \label{sec4}
     
     Since the LHCb Collaboration confirmed the existence of the pentaquark states \cite{LHCb:2015yax}, theorists have shown great interest in predicting the existence of other pentaquark states. Some theoretical groups predict that the $N(1875)$ and $N^{*}(2080)$ are strange partner candidates for $P_{c}$ molecular states \cite{Wu:2023ywu,Ben:2023uev}. In this work, we have studied the $\pi^{-} p\rightarrow\phi n$ scattering process by using the effective Lagrangian approach and discussed the possibility of searching for the hidden-strange pentaquarklike state $N^{*}(2080)$ by $\pi^{-} p$ reactions. The calculation results show that the experimental data of the total cross section and differential cross sections can provide a good constraint on the $t$-channel and $u$-channel contributions in the $\pi^{-} p\rightarrow\phi n$ scattering process. The contribution from the $s$-channel $N^{*}(2080)$ exchange is sensitive to the differential cross section at the forward angle ($\cos\theta = 1$), and a distinct peak appears. However, due to the lack of experimental data and limited accuracy, it is difficult to give a very precise contribution from the $N^{*}(2080)$. Due to the large branching ratio of the $\phi$ decay to $K^{+}K^{-}$, we calculate the Dalitz process of the $\pi^{-}p\rightarrow \phi n \rightarrow K^{+}K^{-}n$ reaction and compare it with the experimental data of $\pi^{-}p\rightarrow K^{+}K^{-}n$. The results show that it is feasible to find the $N^{*}(2080)$ by the $\pi^{-}p\rightarrow \phi n$ process in experiments. It should be mentioned that in our calculation, we consider two cases where the $J^{P}$ quantum number of the $N^{*}(2080)$ is $\frac{1}{2}^{-}$ or $\frac{3}{2}^{-}$, and the results show that the differences of cross sections corresponding to the $N^{*}(2080)$ with different quantum numbers are very small. One of the main reasons is that due to the limited measurement and study of the decay width of $N^{*}$, we can only set the coupling constant related to $N^{*}$ as a free parameter. Therefore, subsequent experimental measurements and theoretical calculations of the internal properties of $N^{*}$ states are highly desirable.

     Currently, the J-PARC \cite{Aoki:2021cqa}, AMBER \cite{Adams:2018pwt} and future HIKE \cite{HIKE:2023ext} and HIAF \cite{Wang:2024} meson beam experiments can measure the meson-nucleus scattering process with high precision. We propose that these experiments make accurate measurements of the $\pi^{-} p\rightarrow\phi n$ process (especially the differential cross section at $\cos\theta=1$), which is important for clarifying the mechanism of the $\pi^{-} p\rightarrow\phi n$ process and determining the contribution of the $N^{*}$ state. At present, the hidden-strange pentaquarklike state has not been  experimentally confirmed, which urgently needs to be clarified through the joint efforts of experiment and theory. It is an effective way to analyze the existence and contribution of hidden-strange pentaquarklike states through the $\pi^{-} p\rightarrow\phi n$ reaction \cite{Wang:2024}. This work is the first step of this research direction, and in the future we will systematically study the contribution of baryon excited states through different $\pi^{-} p$ or $K^{-} p$ scattering, providing the necessary theoretical support for future experiments.

     \section*{ACKNOWLEDGMENTS}

     This work is supported by the National Natural Science Foundation of China under Grants No. 12065014, No. 12047501 and No. 12247101, and by the Natural Science Foundation of Gansu province under Grant No. 22JR5RA266. We acknowledge the West Light Foundation of The Chinese Academy of Sciences, Grant No. 21JR7RA201.
     X.L. is also supported by National Key Research and Development Program of China under Contract No. 2020YFA0406400, the 111 Project under Grant No. B20063, the Fundamental Research Funds for the Central Universities, and the project for top-notch innovative talents of Gansu province.


\begin{thebibliography}{99}
		
		\bibitem{Belle:2003nnu}
		S.~K.~Choi \textit{et al.} [Belle],
		``Observation of a narrow charmonium-like state in exclusive $B^\pm \to K^\pm \pi^+ \pi^- J/\psi$ decays,''
		Phys. Rev. Lett. \textbf{91}, 262001 (2003).
		
		\bibitem{Klempt:2007cp}
		E.~Klempt and A.~Zaitsev,
		``Glueballs, Hybrids, Multiquarks. Experimental facts versus QCD inspired concepts,''
		Phys. Rept. \textbf{454}, 1-202 (2007).
		
		
		
		\bibitem{BESIII:2013ris}
		M.~Ablikim \textit{et al.} [BESIII],
		``Observation of a Charged Charmoniumlike Structure in $e^+e^- \to \pi^+\pi^- J/\psi$ at $\sqrt{s}$ =4.26 GeV,''
		Phys. Rev. Lett. \textbf{110}, 252001 (2013).
		
		
		\bibitem{LHCb:2015yax}
		R.~Aaij \textit{et al.} [LHCb],
		``Observation of $J/\psi p$ Resonances Consistent with Pentaquark States in $\Lambda_b^0 \to J/\psi K^- p$ Decays,''
		Phys. Rev. Lett. \textbf{115}, 072001 (2015).
		
		\bibitem{Chen:2016qju}
		H.~X.~Chen, W.~Chen, X.~Liu and S.~L.~Zhu,
		``The hidden-charm pentaquark and tetraquark states,''
		Phys. Rept. \textbf{639}, 1-121 (2016).
		
		\bibitem{BESIII:2017bua}
		M.~Ablikim \textit{et al.} [BESIII],
		``Determination of the Spin and Parity of the $Z_c(3900)$,''
		Phys. Rev. Lett. \textbf{119}, 072001 (2017).
		
		\bibitem{LHCb:2019kea}
		R.~Aaij \textit{et al.} [LHCb],
		``Observation of a narrow pentaquark state, $P_c(4312)^+$, and of two-peak structure of the $P_c(4450)^+$,''
		Phys. Rev. Lett. \textbf{122}, 222001 (2019).
		
		\bibitem{BESIII:2020qkh}
		M.~Ablikim \textit{et al.} [BESIII],
		``Observation of a Near-Threshold Structure in the $K^+$ Recoil-Mass Spectra in $e^+e^- \rightarrow K^+(D_s^-D^{*0}+D_s^{*-}D^0$),''
		Phys. Rev. Lett. \textbf{126}, 102001 (2021).
		
		\bibitem{LHCb:2020jpq}
		R.~Aaij \textit{et al.} [LHCb],
		``Evidence of a $J/\psi\Lambda$ structure and observation of excited $\Xi^-$ states in the $\Xi^-_b \to J/\psi\Lambda K^-$ decay,''
		Sci. Bull. \textbf{66}, 1278-1287 (2021).
		\bibitem{LHCb:2022ogu}
		R.~Aaij \textit{et al.} [LHCb],
		``Observation of a $J/\psi\Lambda$ Resonance Consistent with a Strange Pentaquark Candidate in $B^{-}\rightarrow J/\psi\Lambda$ Decays,''
		Phys. Rev. Lett. \textbf{131}, 031901 (2023).
		
		
		
		
		
		\bibitem{LHCb:2021vvq}
		R.~Aaij \textit{et al.} [LHCb],
		``Observation of an exotic narrow doubly charmed tetraquark,''
		Nature Phys. \textbf{18}, 751-754 (2022).
		
		\bibitem{Wang:2011rga}
		W.~L.~Wang, F.~Huang, Z.~Y.~Zhang and B.~S.~Zou,
		``$\Sigma_c \bar{D}$ and $\Lambda_c \bar{D}$ states in a chiral quark model,''
		Phys. Rev. C \textbf{84}, 015203 (2011).
		\bibitem{Yang:2011wz}
		Z.~C.~Yang, Z.~F.~Sun, J.~He, X.~Liu and S.~L.~Zhu,
		``The possible hidden-charm molecular baryons composed of anti-charmed meson and charmed baryon,''
		Chin. Phys. C \textbf{36}, 6-13 (2012).
		\bibitem{Wu:2012md}
		J.~J.~Wu, T.~S.~H.~Lee and B.~S.~Zou,
		``Nucleon Resonances with Hidden Charm in Coupled-Channel Models,''
		Phys. Rev. C \textbf{85}, 044002 (2012).
		\bibitem{Li:2014gra}
		X.~Q.~Li and X.~Liu,
		``A possible global group structure for exotic states,''
		Eur. Phys. J. C \textbf{74}, 3198 (2014).
		\bibitem{Chen:2015loa}
		R.~Chen, X.~Liu, X.~Q.~Li and S.~L.~Zhu,
		``Identifying exotic hidden-charm pentaquarks,''
		Phys. Rev. Lett. \textbf{115}, 132002 (2015).
		\bibitem{Karliner:2015ina}
		M.~Karliner and J.~L.~Rosner,
		``New Exotic Meson and Baryon Resonances from Doubly-Heavy Hadronic Molecules,''
		Phys. Rev. Lett. \textbf{115},  122001 (2015).
		
		\bibitem{Wu:2010jy}
		J.~J.~Wu, R.~Molina, E.~Oset and B.~S.~Zou,
		``Prediction of narrow $N^*$ and $\Lambda^*$ resonances with hidden charm above 4 GeV,''
		Phys. Rev. Lett. \textbf{105}, 232001 (2010).
		
		\bibitem{Chen:2019asm}
		R.~Chen, Z.~F.~Sun, X.~Liu and S.~L.~Zhu,
		``Strong LHCb evidence supporting the existence of the hidden-charm molecular pentaquarks,''
		Phys. Rev. D \textbf{100}, 011502 (2019).
		
		\bibitem{Chen:2019bip}
		H.~X.~Chen, W.~Chen and S.~L.~Zhu,
		``Possible interpretations of the $P_c(4312)$, $P_c(4440)$, and $P_c(4457)$,''
		Phys. Rev. D \textbf{100}, 051501 (2019).
		
		\bibitem{He:2019ify}
		J.~He,
		``Study of $P_c(4457)$, $P_c(4440)$, and $P_c(4312)$ in a quasipotential Bethe-Salpeter equation approach,''
		Eur. Phys. J. C \textbf{79}, 393 (2019).
		
		\bibitem{Cheng:2019obk}
		J.~B.~Cheng and Y.~R.~Liu,
		``$P_c(4457)^+$, $P_c(4440)^+$, and $P_c(4312)^+$: molecules or compact pentaquarks?''
		Phys. Rev. D \textbf{100}, 054002 (2019).
		
		\bibitem{Meng:2019ilv}
		L.~Meng, B.~Wang, G.~J.~Wang and S.~L.~Zhu,
		``The hidden charm pentaquark states and $\Sigma_c\bar{D}^{(*)}$ interaction in chiral perturbation theory,''
		Phys. Rev. D \textbf{100}, 014031 (2019).
		
		\bibitem{Liu:2019tjn}
		M.~Z.~Liu, Y.~W.~Pan, F.~Z.~Peng, M.~S\'anchez S\'anchez, L.~S.~Geng, A.~Hosaka and M.~Pavon Valderrama,
		``Emergence of a complete heavy-quark spin symmetry multiplet: seven molecular pentaquarks in light of the latest LHCb analysis,''
		Phys. Rev. Lett. \textbf{122}, 242001 (2019).
		
		\bibitem{Du:2019pij}
		M.~L.~Du, V.~Baru, F.~K.~Guo, C.~Hanhart, U.~G.~Mei\ss{}ner, J.~A.~Oller and Q.~Wang,
		``Interpretation of the LHCb $P_c$ States as Hadronic Molecules and Hints of a Narrow $P_c(4380)$,''
		Phys. Rev. Lett. \textbf{124}, 072001 (2020).
		
		\bibitem{Liu:2023wfo}
		Z.~W.~Liu, J.~X.~Lu, M.~Z.~Liu and L.~S.~Geng,
		``Distinguishing the spins of P$_{c}$(4440) and P$_{c}$(4457) with femtoscopic correlation functions,''
		Phys. Rev. D \textbf{108}, L031503 (2023).
		
		
		
		\bibitem{Chen:2016ryt}
		R.~Chen, J.~He and X.~Liu,
		``Possible strange hidden-charm pentaquarks from $\Sigma_c^{(*)}\bar{D}_s^*$ and $\Xi^{(',*)}_c\bar{D}^*$ interactions,''
		Chin. Phys. C \textbf{41}, 103105 (2017).
		
		\bibitem{Feijoo:2022rxf}
		A.~Feijoo, W.~F.~Wang, C.~W.~Xiao, J.~J.~Wu, E.~Oset, J.~Nieves and B.~S.~Zou,
		``A new look at the Pcs states from a molecular perspective,''
		Phys. Lett. B \textbf{839}, 137760 (2023).
		
		
		\bibitem{Wang:2023ael}
		F.~L.~Wang and X.~Liu,
		``Surveying the mass spectra and the electromagnetic properties of the $\Xi_c^{(\prime,*)}D^{(*)}$ molecular pentaquarks,''
		Phys. Rev. D \textbf{109}, 1 (2024).
		
		\bibitem{Wang:2022nqs}
		F.~L.~Wang, S.~Q.~Luo, H.~Y.~Zhou, Z.~W.~Liu and X.~Liu,
		``Exploring the electromagnetic properties of the $\Xi_c^{(\prime,*)}\bar{D}_{s}^{(*)}$ and $\Omega_c^{(*)}\bar{D}_{s}^{(*)}$ molecular states,''
		Phys. Rev. D \textbf{108}, 034006 (2023).
		
		\bibitem{Zhou:2022gra}
		H.~Y.~Zhou, F.~L.~Wang, Z.~W.~Liu and X.~Liu,
		``Probing the electromagnetic properties of the $\Xi_c^{(*)}D^{(*)}$-type doubly charmed molecular pentaquarks,''
		Phys. Rev. D \textbf{106}, 034034 (2022).
		
		\bibitem{Wang:2022tib}
		F.~L.~Wang, H.~Y.~Zhou, Z.~W.~Liu and X.~Liu,
		``What can we learn from the electromagnetic properties of hidden-charm molecular pentaquarks with single strangeness?,''
		Phys. Rev. D \textbf{106}, 054020 (2022).
		
		\bibitem{Wang:2022mxy}
		F.~L.~Wang and X.~Liu,
		``Emergence of molecular-type characteristic spectrum of hidden-charm pentaquark with strangeness embodied in the $P_{\psi s}^{\Lambda}$(4338) and $P_{cs}$(4459),''
		Phys. Lett. B \textbf{835}, 137583 (2022).
		
		\bibitem{Wang:2023iox}
		F.~L.~Wang and X.~Liu,
		``Higher molecular  $P_{\psi s}^{\Lambda/\Sigma}$ pentaquarks arising from the $\Xi_c^{(\prime,*)}\bar{D}_{1}$/$\Xi_c^{(\prime,*)}\bar{D}_{2}^{(*)}$ interactions,''
		Phys. Rev. D \textbf{108}, 054028 (2023).
		
		\bibitem{Xiao:2019gjd}
		C.~W.~Xiao, J.~Nieves and E.~Oset,
		``Prediction of hidden charm strange molecular baryon states with heavy quark spin symmetry,''
		Phys. Lett. B \textbf{799}, 135051 (2019).
		
		
		\bibitem{Wang:2019nvm}
		B.~Wang, L.~Meng and S.~L.~Zhu,
		``Spectrum of the strange hidden charm molecular pentaquarks in chiral effective field theory,''
		Phys. Rev. D \textbf{101}, 034018 (2020).
		
		
		
		
		
		\bibitem{Wang:2023aob}
		F.~L.~Wang and X.~Liu,
		``New type of doubly charmed molecular pentaquarks containing most strange quarks: Mass spectra, radiative decays, and magnetic moments,''
		Phys. Rev. D \textbf{108}, 074022 (2023).
		
		\bibitem{Wang:2023mdj}
		Z.~Y.~Wang, C.~W.~Xiao, Z.~F.~Sun and X.~Liu,
		``Molecular-type QQss$\bar{s}$ pentaquarks predicted by an extended hidden gauge symmetry approach,''
		Phys. Rev. D \textbf{109}, 034038 (2024).
		
		\bibitem{Wang:2022aga}
		W.~F.~Wang, A.~Feijoo, J.~Song and E.~Oset,
		``Molecular $\Omega_{cc}$,$\Omega_{bb}$, and $\Omega_{bc}$ states,''
		Phys. Rev. D \textbf{106}, 116004 (2022).
		
		\bibitem{Karliner:2022erb}
		M.~Karliner and J.~L.~Rosner,
		``New strange pentaquarks,''
		Phys. Rev. D \textbf{106}, 036024 (2022).
		
		
		
		
		
		\bibitem{Duran:2022xag}
		B.~Duran, Z.~E.~Meziani, S.~Joosten, M.~K.~Jones, S.~Prasad, C.~Peng, W.~Armstrong, H.~Atac, E.~Chudakov and H.~Bhatt, \textit{et al.}
		``Determining the gluonic gravitational form factors of the proton,''
		Nature \textbf{615}, 813-816 (2023).
		
		\bibitem{GlueX:2019mkq}
		A.~Ali \textit{et al.} [GlueX],
		``First Measurement of Near-Threshold J/\ensuremath{\psi} Exclusive Photoproduction off the Proton,''
		Phys. Rev. Lett. \textbf{123}, 072001 (2019).
		
		\bibitem{GlueX:2023pev}
		S.~Adhikari \textit{et al.} [GlueX],
		``Measurement of the J/$\psi $ photoproduction cross section over the full near-threshold kinematic region,''
		Phys. Rev. C \textbf{108}, 025201 (2023).
		
		\bibitem{Gao:2017hya}
		H.~Gao, H.~Huang, T.~Liu, J.~Ping, F.~Wang and Z.~Zhao,
		``Search for a hidden strange baryon-meson bound state from \ensuremath{\phi} production in a nuclear medium,''
		Phys. Rev. C \textbf{95}, 055202 (2017).
		
		
		\bibitem{He:2003vi}
		J.~He and Y.~B.~Dong,
		``Negative parity $N^{*}$ resonances in an extended GBE,''
		Nucl. Phys. A \textbf{725}, 201-210 (2003).
		
		\bibitem{He:2003nss}
		J.~He and D.~Yu-bing,
		``Test of one pion exchange and one gluon exchange through mixing angles of negative parity $N^{*}$ resonances in electromagnetic transitions,''
		Phys. Rev. D \textbf{68}, 017502 (2003).
		
		\bibitem{He:2015yva}
		J.~He,
		``Internal structures of the nucleon resonances N(1875) and N(2120),''
		Phys. Rev. C \textbf{91}, 018201 (2015).
		
		\bibitem{Yuan:2010cfa}
		S.~G.~Yuan, C.~S.~An and J.~He,
		``Contributions of $qqqq\bar q$ components to axial charges of proton and $N(1440)$,''
		Commun. Theor. Phys. \textbf{54}, 697-700 (2010).
		
		\bibitem{An:2018vmk}
		C.~S.~An, J.~J.~Xie and G.~Li,
		``Decay patterns of low-lying $N_{s\bar{s}}$ states to the strangeness channels,''
		Phys. Rev. C \textbf{98}, 045201 (2018).
		
		
		
		\bibitem{Wu:2011zzm}
		J.~J.~Wu, R.~Molina, E.~Oset and B.~S.~Zou,
		``Prediction of narrow $N^{*}$ and $\Lambda^{*}$ with hidden charm,''
		AIP Conf. Proc. \textbf{1374}, 557-560 (2011).
		
		\bibitem{Zou:2005xy}
		B.~S.~Zou and D.~O.~Riska,
		``The $s\bar{s}$ component of the proton and the strangeness magnetic moment,''
		Phys. Rev. Lett. \textbf{95}, 072001 (2005).
		
		
		\bibitem{An:2009uv}
		C.~S.~An and B.~S.~Zou,
		``Strong decays of $N^{*}(1535)$ in an extended chiral quark model,''
		Sci. China G \textbf{52}, 1452-1457 (2009).
		
		
		\bibitem{Cao:2009ea}
		X.~Cao, J.~J.~Xie, B.~S.~Zou and H.~S.~Xu,
		``Evidence of N$^{*}$(1535) resonance contribution in the $pn\rightarrow d\phi$ reaction,''
		Phys. Rev. C \textbf{80}, 025203 (2009).
		
		\bibitem{Ben:2023uev}
		D.~Ben, A.~C.~Wang, F.~Huang and B.~S.~Zou,
		``Effects of $N(2080)3/2^{-}$ and $N(2270)3/2^{-}$ molecules on $K^*\Sigma$ photoproduction,''
		Phys. Rev. C \textbf{108}, 065201 (2023).
		
		\bibitem{Xie:2010yk}
		J.~J.~Xie and J.~Nieves,
		``The role of the $N^*(2080)$ resonance in the $\vec{\gamma} p \to K^+ \Lambda(1520)$ reaction,''
		Phys. Rev. C \textbf{82}, 045205 (2010).
		
		
		\bibitem{Lin:2018kcc}
		Y.~H.~Lin, C.~W.~Shen and B.~S.~Zou,
		``Decay behavior of the strange and beauty partners of $P_c$ hadronic molecules,''
		Nucl. Phys. A \textbf{980}, 21-31 (2018).
		
		\bibitem{He:2017aps}
		J.~He,
		``Nucleon resonances $N(1875)$ and $N(2100)$ as strange partners of LHCb pentaquarks,''
		Phys. Rev. D \textbf{95}, 074031 (2017).
		
		\bibitem{Workman:2022ynf}
		R.~L.~Workman \textit{et al.} [Particle Data Group],
		``Review of Particle Physics,''
		PTEP \textbf{2022}, 083C01 (2022).
		
		\bibitem{Wu:2023ywu}
		S.~M.~Wu, F.~Wang and B.~S.~Zou,
		``Strange molecular partners of Pc states in the \ensuremath{\gamma}p\textrightarrow{}\ensuremath{\phi}p reaction,''
		Phys. Rev. C \textbf{108}, 045201 (2023).
		
		
		
		\bibitem{Cheng:2016ddp}
		C.~Cheng and X.~Y.~Wang,
		``The production of neutral $N^*(11052)$ resonance with hidden beauty from $\pi^-p$ scattering,''
		Adv. High Energy Phys. \textbf{2017}, 9398732 (2017).
		
		\bibitem{Wang:2017qcw}
		X.~Y.~Wang and J.~He,
		``Investigation of pion-induced $f_1(1285)$ production off a nucleon target within an interpolating Reggeized approach,''
		Phys. Rev. D \textbf{96},  034017 (2017).
		
		\bibitem{Wang:2019zaw}
		X.~Y.~Wang, J.~He and X.~Chen,
		``Systematic study of the production of hidden-bottom pentaquarks via $\gamma p$ and $\pi ^{-}p$ scatterings,''
		Phys. Rev. D \textbf{101}, 034032 (2020).
		
		\bibitem{Xie:2013db}
		J.~J.~Xie and B.~C.~Liu,
		``Role of the $N^*(2080)$ in $pp \to pK^+ \Lambda(1520)$ and $\pi^- p \to K^0 \Lambda(1520)$ reactions,''
		Phys. Rev. C \textbf{87}, 045210 (2013). 
		
		\bibitem{Wang:2019dsi}
		X.~Y.~Wang, J.~He, X.~R.~Chen, Q.~Wang and X.~Zhu,
		``Pion-induced production of hidden-charm pentaquarks $P_{c}(4312),P_{c}(4440)$, and $P_{c}(4457)$,''
		Phys. Lett. B \textbf{797}, 134862 (2019).
		
		\bibitem{Clymton:2022qlr}
		S.~Clymton, H.~J.~Kim and H.~C.~Kim,
		``The effect of hidden-charm strange pentaquarks $p_{cs}$ on the $K^- p \rightarrow J/\psi \Lambda$~ reaction,''
		Rev. Mex. Fis. Suppl. \textbf{3}, 0308040 (2022).
		
		
		\bibitem{Lu:2015pva}
		Q.~F.~L\"u and D.~M.~Li,
		``Near-threshold \ensuremath{\eta} production in pp collisions,''
		Chin. Phys. C \textbf{39}, 113104 (2015).
		
		
		\bibitem{Wang:2019krd}
		X.~Y.~Wang, X.~R.~Chen and J.~He,
		``Possibility to study pentaquark states $P_{c}(4312), P_{c}(4440)$, and $P_{c}(4457)$ in $\gamma p\rightarrow J/\psi p$ reaction,''
		Phys. Rev. D \textbf{99}, 114007 (2019).
		
		\bibitem{Kim:2016cxr}
		S.~H.~Kim, H.~C.~Kim and A.~Hosaka,
		``Heavy pentaquark states $P_c(4380)$ and $P_c(4450)$ in the $J/\psi$ production induced by pion beams off the nucleon,''
		Phys. Lett. B \textbf{763}, 358-364 (2016).
		
		
		
		\bibitem{Ryu:2012tw}
		H.~Y.~Ryu, A.~I.~Titov, A.~Hosaka and H.~C.~Kim,
		``$\phi$ photoprodution with coupled-channel effects,''
		PTEP \textbf{2014}, 023D03 (2014).
		
		
		
		
		\bibitem{Stoks:1999bz}
		V.~G.~J.~Stoks and T.~A.~Rijken,
		``Soft core baryon baryon potentials for the complete baryon octet,''
		Phys. Rev. C \textbf{59}, 3009-3020 (1999).
		
		\bibitem{Kim:2021adl}
		S.~H.~Kim, T.~S.~H.~Lee, S.~i.~Nam and Y.~Oh,
		``Dynamical model of \ensuremath{\phi} meson photoproduction on the nucleon and $^{4}$He,''
		Phys. Rev. C \textbf{104}, 045202 (2021).
		
		\bibitem{Lu:2015fva}
		Q.~F.~L\"u, X.~Y.~Wang, J.~J.~Xie, X.~R.~Chen and Y.~B.~Dong,
		``Neutral hidden charm pentaquark states $P_c^0(4380)$ and $P_c^0(4450)$ in $\pi^-p \to J/\psi n$ reaction,''
		Phys. Rev. D \textbf{93}, 034009 (2016).
		
		
		
		
		\bibitem{Wang:2015xwa}
		X.~Y.~Wang and X.~R.~Chen,
		``Production of the superheavy baryon $\Lambda_{c\bar{c}}^{*}$ (4209) in kaon-induced reaction,''
		Eur. Phys. J. A \textbf{51}, 85 (2015).
		
		
		
		
		\bibitem{Wang:2023lia}
		X.~Y.~Wang, H.~F.~Zhou and X.~Liu,
		``Exploring kaon induced reactions for unraveling the nature of the scalar meson $a_0 (1817)$,''
		Phys. Rev. D \textbf{108}, 034015 (2023).
		
		\bibitem{Wang:2015kia}
		X.~Y.~Wang and A.~Guskov,
		``Photoproduction of $a_{2}$(1320) in a Regge model,''
		Phys. Rev. D \textbf{93}, 074016 (2016).
		
		\bibitem{Liu:2021ojf}
		J.~Liu, D.~Y.~Chen and J.~He,
		``Double exotic state productions in pion and kaon induced reactions,''
		Eur. Phys. J. C \textbf{81}, 965 (2021).

		\bibitem{Courant:1977rk}
		H.~Courant, Y.~I.~Makdisi, M.~L.~Marshak, E.~A.~Peterson, K.~Ruddick and J.~Smith-Kintner,
		``Phi Production in pi- p Collisions Near Threshold,''
		Phys. Rev. D \textbf{16}, 1-6 (1977).

\bibitem{Dahl:1967pg}
O.~I.~Dahl, L.~M.~Hardy, R.~I.~Hess, J.~Kirz and D.~H.~Miller,
``Strange-particle production in pi- p interactions from 1.5 to 4.2 BeV/c. 1. Three-and-more-body final states,''
Phys. Rev. \textbf{163}, 1377-1429 (1967).

    \bibitem{Kim:2017nxg}
    S.~H.~Kim, S.~i.~Nam, D.~Jido and H.~C.~Kim,
    ``Photoproduction of $\Lambda (1405)$ with the $N^*$ and $t$-channel Regge contributions,''
    Phys. Rev. D \textbf{96}, 014003 (2017).

\bibitem{Boyd:1968pg}
J.~H.~Boyd, A.~R.~Erwin, W.~D.~Walker and E.~West,
``Study of $\pi^{-}p\rightarrow \omega n, \phi n$ at 2.10 BeV/c,''
Phys. Rev. \textbf{166}, 1458-1468 (1968).

\bibitem{Miller:1965pg}
D.~H.~Miller, A.~Z.~Kovacs, R.~Mcllwain, T.~R.~Palfrey and G.~W.~Tautfest,
``Strange-Particle Profuction in 2.7-GeV/c $\pi^{-}p$ Interactions,''
Phys. Rev. \textbf{140}, B360-365 (1965).

\bibitem{Goussu:1966pg}
O.~Goussu, G.~Smadja, G.~Kayas, ``Production de particules étranges dans les réactions $\pi^{-}p$ à 2.75 GeV/c ,''. Nuovo Cimento A (1965-1970) 47, 383–399 (1967). 
  
		\bibitem{Aoki:2021cqa}
		K.~Aoki, H.~Fujioka, T.~Gogami, Y.~Hidaka, E.~Hiyama, R.~Honda, A.~Hosaka, Y.~Ichikawa, M.~Ieiri and M.~Isaka, \textit{et al.}
		``Extension of the J-PARC Hadron Experimental Facility: Third White Paper,''
		[arXiv:2110.04462 [nucl-ex]].
		
		\bibitem{Adams:2018pwt}
		B.~Adams, C.~A.~Aidala, R.~Akhunzyanov, G.~D.~Alexeev, M.~G.~Alexeev, A.~Amoroso, V.~Andrieux, N.~V.~Anfimov, V.~Anosov and A.~Antoshkin, \textit{et al.}
		``Letter of Intent: A New QCD facility at the M2 beam line of the CERN SPS (COMPASS++/AMBER),''
		[arXiv:1808.00848 [hep-ex]].
		
		\bibitem{HIKE:2023ext}
		M.~U.~Ashraf \textit{et al.} [HIKE],
		``High Intensity Kaon Experiments (HIKE) at the CERN SPS Proposal for Phases 1 and 2,''
		[arXiv:2311.08231 [hep-ex]].
		
		\bibitem{Wang:2024}
		X.~Y.~Wang and X.~Liu,
		``Development status and future plan of meson beam experiments at home and abroad,''
		ChinaXiv:202402.00079.
		
		
	\end{thebibliography}
\end{document}